%% file: eccv2020submission.tex
\documentclass[runningheads]{llncs}
\usepackage{graphicx}
\usepackage{comment}
\usepackage{amsmath,amssymb} 
\usepackage{color}

\usepackage{epsfig}
\usepackage{amsfonts,bm}

\usepackage{makecell}
\usepackage{xcolor}
\usepackage{multirow}
\usepackage{fixltx2e}
\usepackage{pifont}
\usepackage{mmstyle}
\usepackage{hyperref}

\newcommand{\xg}[1]{{\color{blue}(xg: {#1})}} 

\usepackage[width=122mm,left=12mm,paperwidth=146mm,height=193mm,top=12mm,paperheight=217mm]{geometry}

\begin{document}
	\pagestyle{headings}
	\mainmatter
	\def\ECCVSubNumber{3265}  
	
	\title{Exploiting Deep Generative Prior for \\ Versatile Image Restoration and Manipulation \vspace{-0.1cm} } 

	\titlerunning{Deep Generative Prior}
	%
	
	\author{Xingang Pan\inst{1} \and \
	Xiaohang Zhan\inst{1} \and \
	Bo Dai\inst{1} \and \\
	Dahua Lin\inst{1} \and \
	Chen Change Loy\inst{2} \and \
	Ping Luo\inst{3}}
	\authorrunning{X. Pan et al.}
	%
	\institute{The Chinese University of Hong Kong \\
		\email{\{px117,zx017,bdai,dhlin\}@ie.cuhk.edu.hk}\\
		\and Nanyang Technological University ~~ \inst{3} The University of Hong Kong \\
		~~ \email{ccloy@ntu.edu.sg}  \qquad\qquad\quad \email{pluo@cs.hku.hk} }
	\maketitle
	
	\input{./sections/abstract.tex}

	\input{./sections/introduction.tex}

	\input{./sections/relatedwork.tex}
	\input{./sections/dgp.tex}

\input{./sections/experiments.tex}
	\input{./sections/conclusion.tex}

	%
	%
	\bibliographystyle{splncs04}
	\bibliography{egbib}
	
	\input{./sections/appendix.tex}
\end{document}

%% file: sections/abstract.tex

\begin{abstract}
	
	
	\vspace{-0.3cm}
	
	Learning a good image prior is a long-term goal for image restoration and manipulation.
	While existing methods like deep image prior (DIP) capture low-level image statistics, there are still gaps toward an image prior that captures rich image semantics including color, spatial coherence, textures, and high-level concepts.
	This work presents an effective way to exploit the image prior captured by a generative adversarial network (GAN) trained on large-scale natural images. As shown in Fig.~\ref{fig:intro}, the deep generative prior (DGP) provides compelling results to restore missing semantics, \eg, color, patch, resolution, of various degraded images. It also enables diverse image manipulation including random jittering, image morphing, and category transfer. Such highly flexible restoration and manipulation are made possible through relaxing the assumption of existing GAN-inversion methods, which tend to fix the generator. Notably, we allow the generator to be fine-tuned on-the-fly in a progressive manner regularized by feature distance obtained by the discriminator in GAN. We show that these easy-to-implement and practical changes help preserve the reconstruction to remain in the manifold of nature image, and thus lead to more precise and faithful reconstruction for real images. Code is available at \href{https://github.com/XingangPan/deep-generative-prior}{https://github.com/XingangPan/deep-generative-prior}. 

\end{abstract}

%% file: sections/introduction.tex

\section{Introduction}

Learning image prior models is important to solve various tasks of image restoration and manipulation, such as \textit{image colorization}~\cite{larsson2016learning,zhang2016colorful}, 
\textit{image inpainting}~\cite{yeh2017semantic},
\textit{super-resolution}~\cite{dong2015image,ledig2017photo},
and \textit{adversarial defense}~\cite{samangouei2018defense}. 
In the past decades, many image priors~\cite{roth2005fields,zhu1997prior,geman1984stochastic,he2010single,rudin1992nonlinear} have been proposed to capture certain statistics of natural images.  
Despite their successes, these priors often serve a dedicated purpose.
For instance, markov random field~\cite{roth2005fields,zhu1997prior,geman1984stochastic} is often used to model the correlation among neighboring pixels,
while dark channel prior~\cite{he2010single} and total variation~\cite{rudin1992nonlinear} are developed for dehazing and denoising respectively.

%

There is a surge of interest to seek for more general priors that capture richer statistics of images through deep learning models.
For instance, the seminal work on deep image prior (DIP)~\cite{ulyanov2018deep} showed that the structure of a randomly initialized Convolutional Neural Network (CNN) implicitly captures texture-level image prior, thus can be used for restoration by fine-tuning it to reconstruct a corrupted image.
SinGAN~\cite{shaham2019singan} further shows that a randomly-initialized generative adversarial network (GAN) model is able to capture rich patch statistics after training from a single image.
These priors have shown impressive results on some low-level image restoration and manipulation tasks like super-resolution and harmonizing.
In both the representative works, the CNN and GAN are trained from a single image of interest from scratch.

\begin{figure*}[t]
	\centering
	\includegraphics[width=\linewidth]{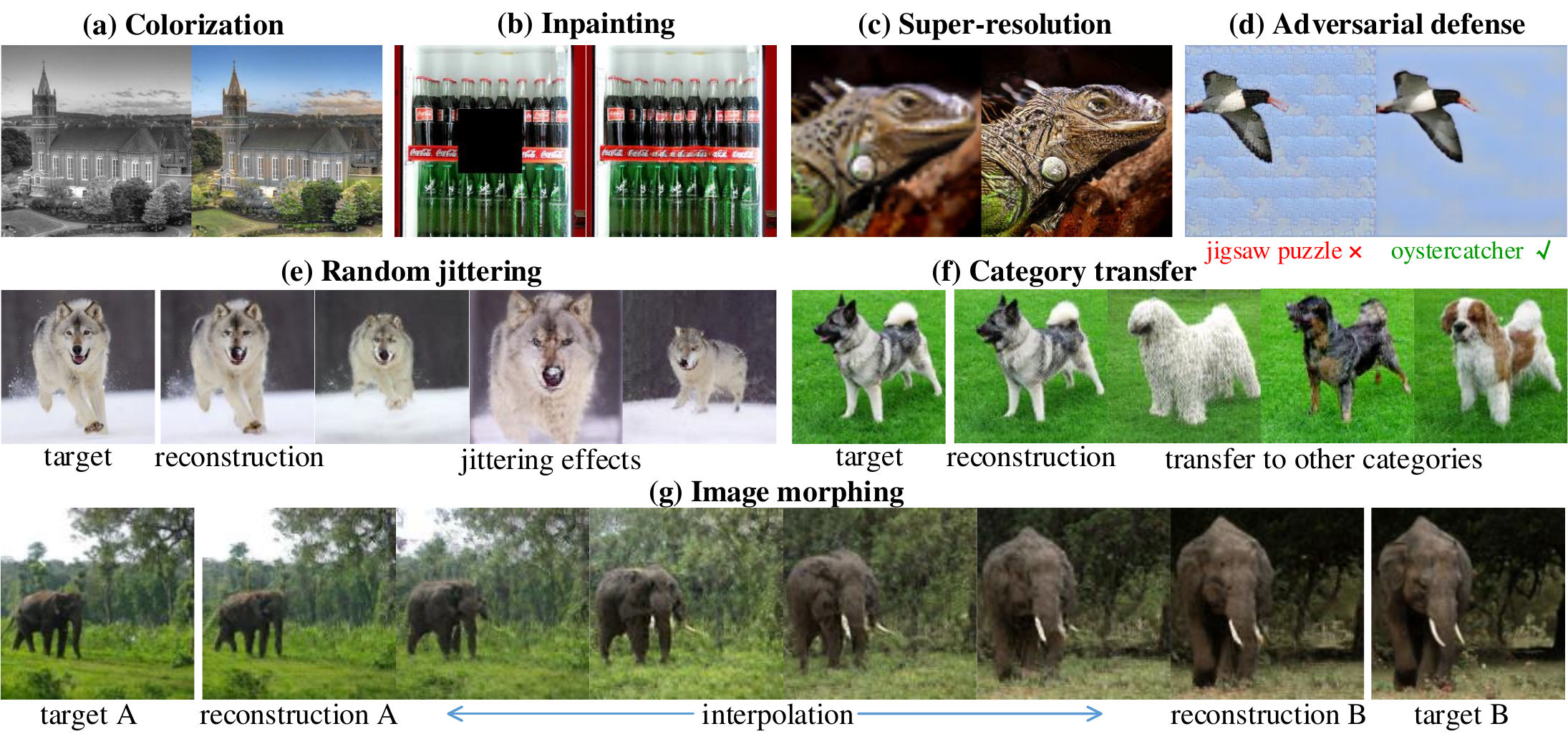}
	\vspace{-0.7cm}
	\caption{\hspace{-1pt}These image restoration(a)(b)(c)(d)\hspace{-0.5pt} and manipulation(e)(f)(g)\hspace{-0.5pt} effects are achieved by leveraging the rich generative prior of a GAN. The GAN does not see these images during training }
	\vspace{-0.2cm}
	\label{fig:intro}
\end{figure*}

In this study, we are interested to go one step further, examining how we could leverage a GAN~\cite{goodfellow2014generative} trained on large-scale natural images for richer priors beyond a single image.
%
GAN is a good approximator for natural image manifold. 
By learning from large image datasets, it captures rich knowledge on natural images including color, spatial coherence, textures, and high-level concepts, which are useful for broader image restoration and manipulation effects.
Specifically, we take a collapsed image (\eg, gray-scale image) as a partial observation of the original natural image, and reconstruct it in the observation space (\eg, gray-scale space) with the GAN, the image prior of the GAN would tend to restore the missing semantics (\eg, color) in a faithful way to match natural images.
%
Despite its enormous potentials, it remains a challenging task to exploit a GAN as a prior for general image restoration and manipulation. 
The key challenge lies in the needs in coping with arbitrary images from different tasks with distinctly different natures. The reconstruction also needs to produce sharp and faithful images obeying the natural image manifold.



\begin{figure}[t!]
	\centering
	\includegraphics[width=\linewidth]{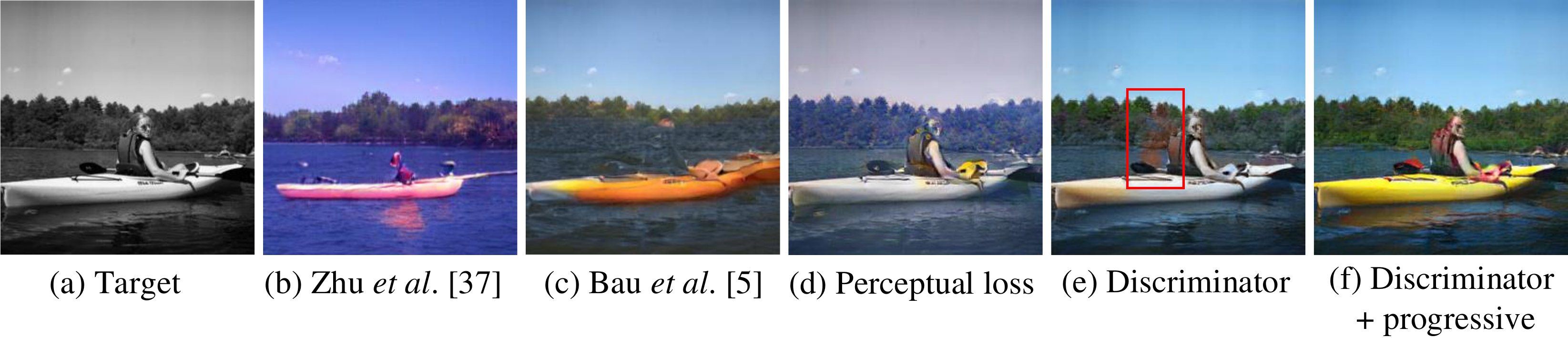}
	\vspace{-0.7cm}
	\caption{Comparison of various methods in reconstructing a gray image under the gray-scale observation space using a GAN. Conventional GAN-inversion strategies like (b)\cite{zhu2016generative} and (c)\cite{bau2019seeing} produce imprecise reconstruction for the existing semantics. In this work, we relax the generator so that it can be fine-tuned on-the-fly, achieving more accurate reconstruction as in (d)(e)(f), of which optimization is based on (d) VGG perceptual loss, (e) discriminator feature matching loss, and (f)  combined with progressive reconstruction, respectively. We highlight that discriminator is important to preserve the generative prior so as to achieve better restoration for the missing information (\ie, color). The proposed progressive strategy eliminates the `information lingering' artifacts as in the red box in (e) }
	\label{fig:gan_inversion}
	\vspace{-0.2cm}
\end{figure}

An appealing option for our problem is GAN-inversion~\cite{zhu2016generative,creswell2018inverting,Albright2019CVPRWorkshops,bau2019seeing}.
Existing GAN-inversion methods typically reconstruct a target image by optimizing over the latent vector, \ie, \(\vz^* = \argmin_{\vz \in \Rbb^d} \cL(\vx, G(\vz; \vtheta))\), where $\vx$ is the target image, $G$ is a fixed generator,
 $\vz$ and $\vtheta$ are the latent vector and generator parameters, respectively.
In practice, we found that this strategy fails in dealing with complex real-world images. In particular, it often results in mismatched reconstructions, whose details (\eg, objects, texture, and background) appear inconsistent with the original images, as Fig.~\ref{fig:gan_inversion} (b)(c) show.
On one hand, existing GAN-inversion methods still suffer from the issues of mode collapse and limited generator capacity, affecting their capability in capturing the desired data manifold.
On the other hand, perhaps a more crucial limitation is that when a generator is fixed, the GAN is inevitably limited by the training distribution and its inversion cannot faithfully reconstruct unseen and complex images.
It is infeasible to carry such assumptions while using a GAN as prior for general image restoration and manipulation.


Despite the gap between the approximated manifold and the real one,
the GAN generator still captures rich statistics of natural images.
In order to make use of these statistics while avoiding the aforementioned limitation,
in this paper we present a relaxed and more practical reconstruction formulation for mining the priors in GAN.
Our first reformulation is to allow the generator parameters to be fine-tuned on the target image on-the-fly, 
\ie, \(\vtheta^*, \vz^* = \argmin_{\vtheta, \vz} \cL(\vx, G(\vz; \vtheta))\). This lifts the constraint of confining the reconstruction within the training distribution. 
Relaxing the assumption with fine-tuning, however, is still not sufficient to ensure good reconstruction quality for arbitrary target images. 
We found that fine-tuning using a standard loss such as perceptual loss~\cite{johnson2016perceptual} or mean squared error (MSE) in DIP could risk wiping out the originally rich priors. Consequently, the reconstruction may become increasingly unnatural during the reconstruction of a degraded image.
Fig.~\ref{fig:gan_inversion}(d) shows an example, suggesting that a new loss and reconstruction strategy is needed. 

Thus, in our second reformulation, we devise an effective reconstruction strategy that consists of two components: 

\noindent
1) \textit{Feature matching loss from the coupled discriminator} - we make full use of the discriminator of a trained GAN to regularize the reconstruction. 
Note that during training, the generator is optimized to mimic massive natural images via gradients provided by the discriminator. 
It is reasonable to still adopt the discriminator in guiding the generator to match a single image as the discriminator preserves the original parameter structure of the generator better than other distance metrics. 
Thus deriving a feature matching loss from the discriminator can help maintain the reconstruction to remain in the natural image space. Although the feature matching loss is not new in the literature~\cite{wang2018high}, its significance to GAN reconstruction has not been investigated before. 

\noindent
2) \textit{Progressive reconstruction} - we observe that a joint fine-tuning of all parameters of the generator could lead to \textit{`information lingering'},  where missing semantics (\eg, color) do not naturally change along with the content when reconstructing a degraded image.
This is because the deep layers of the generator start to match the low-level textures before the high-level configurations are aligned.
To address this issue, we propose a progressive reconstruction strategy that fine-tunes the generator gradually from the shallowest layers to the deepest layers.
This allows the reconstruction to start with matching high-level configurations and gradually shift its focus on low-level details.
%

Thanks to the proposed techniques that enable faithful reconstruction while maintaining the generator prior,
our approach, which we name as Deep Generative Prior (DGP), generalizes well to various kinds of image restoration and manipulation tasks, despite that our method is not specially designed for each task.
When reconstructing a corrupted image in a task-dependent observation space, DGP tends to restore the missing information,
while keeping existing semantic information unchanged. As shown in
Fig.~\ref{fig:intro} (a)(b)(c), color, missing patches, and details of the given images are well restored, respectively.
As illustrated in Fig.~\ref{fig:intro} (e)(f),
we can manipulate the content of an image by tweaking the latent vector or category condition of the generator.
Fig.~\ref{fig:intro} (g) shows that image morphing is possible by interpolating between the parameters of two fine-tuned generators and the corresponding latent vectors of these images. 
To our knowledge, 
it is the first time these jittering and morphing effects are achieved on a dataset with complex images like ImageNet~\cite{deng2009imagenet}. We show more interesting examples in the experiments and Appendix.



%% file: sections/relatedwork.tex

\section{Related Work}

\noindent\textbf{Image Prior.}
Image priors that describe various statistics of natural images have been widely adopted in computer vision,
including markov random fields \cite{roth2005fields,zhu1997prior,geman1984stochastic},
dark channel prior \cite{he2010single},
and total variation regularizer \cite{rudin1992nonlinear}.
Recently, the work of deep image prior (DIP)~\cite{ulyanov2018deep} shows that image statistics are implicitly captured by the structure of CNN,
which is also a kind of prior, and could be used to restore corrupted images.
%
SinGAN~\cite{shaham2019singan} fine-tunes a randomly initialized GAN on patches of a single image,
achieving various image editing or restoration effects.
As DIP and SinGAN are trained from scratch, they have limited access to image statistics beyond the input image,
which restrains their applicability in tasks such as image colorization.
There are also other deep priors developed for low-level restoration tasks like deep denoiser prior~\cite{zhang2017learning,bigdeli2017deep} and TNRD~\cite{chen2016trainable}, but competing with them is not our goal.
Instead, our goal is to study and exploit the prior that is captured in GAN for versatile restoration as well as manipulation tasks.
Existing attempts that use a pre-trained GAN as a source of image statistics include \cite{bau2019semantic} and \cite{hussein2019image},
which respectively applies to image manipulation, \eg,~editing partial areas of an image, and image restoration, \eg,~compressed sensing and super-resolution for human faces.
As we will show in our experiments, by using a discriminator based distance metric and a progressive fine-tuning strategy,
DGP can better preserve image statistics learned by the GAN and thus allows richer restoration and manipulation effects.

Recently, a concurrent work of multi-code GAN prior~\cite{gu2020image} also conducts image processing by solving the GAN-inversion problem.
It uses multiple latent vectors to reconstruct the target image and keeps the generator fixed, while our method makes the generator image-adaptive by allowing it to be fine-tuned on-the-fly.

\vspace{3pt}
\noindent\textbf{Image Restoration and Manipulation.}
In this paper we demonstrate the effect of applying DGP to multiple tasks of image processing,
including image colorization~\cite{larsson2016learning}, 
image inpainting~\cite{yeh2017semantic}, 
super-resolution~\cite{dong2015image,ledig2017photo}, adversarial defence~\cite{samangouei2018defense}, and semantic manipulation~\cite{zhu2016generative,zhu2017unpaired,choi2018stargan}.
While many task-specific models and loss functions have been proposed to pursue a better performance on a specific restoration task \cite{larsson2016learning,zhang2016colorful,yeh2017semantic,dong2015image,ledig2017photo,samangouei2018defense},
there are also works that apply GAN and design task-specific pipelines to achieve various image manipulation effects \cite{zhu2017unpaired,choi2018stargan,wang2018high,bau2019semantic,shen2019interpreting,yang2019semantic}, such as CycleGAN~\cite{zhu2017unpaired} and StarGAN~\cite{choi2018stargan}.
In this work we are more interested in uncovering the potential of exploiting the GAN prior as a task-agnostic solution, where we propose several techniques to achieve this goal.
Moreover, as shown in Fig.~\ref{fig:intro}(e)(g),
with an improved reconstruction process we successfully achieve image jittering and morphing on ImageNet,
while previous methods are insufficient to handle these effects on such complex data.

\vspace{3pt}
\noindent\textbf{GAN-Inversion.}
As mentioned in Sec.\ref{sec:dgp}, a straightforward way to utilize generative prior is conducting image reconstruction based on GAN-inversion.
GAN-inversion aims at finding a vector in the latent space that best reconstructs a given image, where the GAN generator is fixed.
Previous attempts either 
optimize the latent vector directly via gradient back-propagation~\cite{creswell2018inverting,Albright2019CVPRWorkshops} 
or leverage an additional encoder mapping images to latent vectors~\cite{zhu2016generative,donahue2017adversarial}.
A more recent approach \cite{bau2019seeing} proposes to add small perturbations to shallow blocks of the generator to ease the inversion task.
While these methods could handle datasets with limited complexities or synthetic images sampled by the GAN itself,
we empirically found in our experiments they may produce imprecise reconstructions for complex real scenes, \eg,~images in the ImageNet~\cite{deng2009imagenet}.
Recently, the work of StyleGAN \cite{karras2019style} enables a new way for GAN-inversion by operating in intermediate latent spaces \cite{abdal2019image2stylegan},
but noticeable mismatches are still observed and the inversion for vanilla GAN (\eg, BigGAN~\cite{brock2018large}) is still challenging.
In this paper, instead of directly applying standard GAN-inversion,
we devise a more practical way to reconstruct a given image using the generative prior,
which is shown to achieve better reconstruction results.

%% file: sections/dgp.tex

\section{Method}
\label{sec:dgp}


We first provide some preliminaries on DIP and GAN before discussing how we exploit DGP for image restoration and manipulation.

\vspace{0.1cm}
\noindent\textbf{Deep Image Prior.}
Ulyanov \etal\cite{ulyanov2018deep} show that image statistics are implicitly captured by the structure of CNN. These statistics can be seen as a kind of image prior, which can be exploited in various image restoration tasks 
by tuning a randomly initialized CNN on the degraded image:
$\vtheta^* = \argmin_{\vtheta} E(\hat{\vx}, f(\vz;\vtheta)), \vx^* = f(\vz;\vtheta^*),$
where $E$ is a task-dependent distance metric, 
$\vz$ is a randomly chosen latent vector, and $f$ is a CNN with $\vtheta$ being its parameters.
$\hat{\vx}$ and $\vx^*$ are the degraded image and restored image respectively.
One limitation of DIP is that the restoration process mainly resorts to existing statistics in the input image,
it is thus infeasible to apply DIP on tasks that require more general statistics, such as image colorization~\cite{larsson2016learning} and manipulation~\cite{zhu2016generative}.

\vspace{2pt}
\noindent\textbf{Generative Adversarial Networks (GANs).}
GANs are widely used for modeling complex data such as natural images~\cite{goodfellow2014generative,xiangli2020real,dai2017towards,karras2019style}.
In GAN, the underlying manifold of natural images is approximated by the combination of a parametric generator $G$ and a prior latent space $\cZ$, 
so that an image can be generated by sampling a latent vector $\vz$ from $\cZ$ and applying $G$ as $G(\vz)$.
GAN jointly trains $G$ with a parametric discriminator $D$ in an adversarial manner, 
where $D$ is supposed to distinguish generated images from real ones.
Although extensive efforts have been made to improve the power of GAN,
there inevitably exists a gap between GAN's approximated manifold and the actual one,
due to issues such as insufficient capacity and mode collapse.



\subsection{Deep Generative Prior}

Suppose $\hat{\vx}$ is obtained via $\hat{\vx} = \phi(\vx)$, where $\vx$ is the original natural image and $\phi$ is a degradation transform.
\eg,~$\phi$ could be a graying transform that turns $\vx$ into a grayscale image.
Many tasks of image restoration can be regarded as recovering $\vx$ given $\hat{\vx}$.
A common practice is learning a mapping from $\hat{\vx}$ to $\vx$, which often requires task-specific training for different $\phi$s.
Alternatively, we can also employ statistics of $\vx$ stored in some prior, and search in the space of $\vx$ for an optimal $\vx$ that best matches $\hat{\vx}$, 
viewing $\hat{\vx}$ as partial observations of $\vx$.

While various priors have been proposed~\cite{roth2005fields,ulyanov2018deep,shaham2019singan} in the second line of research,
in this paper we are interested in studying a more generic image prior,
\ie,~a GAN generator trained on large-scale natural images for image synthesis.
Specifically, a straightforward realization is a reconstruction process based on GAN-inversion, 
which optimizes the following objective:
\begin{align}
\vz^* &= \argmin_{\vz \in \Rbb^d} E(\hat{\vx}, G(\vz; \vtheta)), \qquad \vx^* = G(\vz^*; \vtheta), \label{eq:reconstruction} \\
	  &= \argmin_{\vz \in \Rbb^d} \cL(\hat{\vx}, \phi(G(\vz; \vtheta))),  \nonumber
\end{align}
where $\cL$ is a distance metric such as the L2 distance, 
\textit{G} is a GAN generator parameterized by $\vtheta$ and trained on natural images.
Ideally, if $G$ is sufficiently powerful that the data manifold of natural images is well captured in $G$,
the above objective will drag $\vz$ in the latent space 
and locate the optimal natural image $\vx^* = G(\vz^*;\vtheta)$,
which contains the missing semantics of $\hat{\vx}$ and matches $\hat{\vx}$ under $\phi$.
For example, if $\phi$ is a graying transform, $\vx^*$ will be an image with a natural color configuration subject to $\phi(\vx^*) = \hat{\vx}$.
However, in practice it is not always the case.

As the GAN generator is fixed in Eq.\eqref{eq:reconstruction} and its improved versions, \eg,~adding an extra encoder~\cite{zhu2016generative,donahue2017adversarial},
these reconstruction methods based on the standard GAN-inversion suffer from an intrinsic limitation, \ie, there is a gap between the approximated manifold of natural images and the actual one.
On one hand, due to issues including mode collapse and insufficient capacity,
the GAN generator cannot perfectly grasp the training manifold represented by a dataset of natural images.
On the other hand, the training manifold itself is also an approximation of the actual one.
Such two levels of approximations inevitably lead to a gap.
Consequently,  
a sub-optimal $\vx^*$ is often retrieved, which often contains significant mismatches to $\hat{\vx}$,
especially when the original image $\vx$ is a complex image, \eg,~ImageNet \cite{deng2009imagenet} images, or an image located outside the training manifold.
See Fig.~\ref{fig:gan_inversion} and existing literature~\cite{bau2019seeing,donahue2017adversarial} for an illustration.

\vspace{2pt}
\noindent\textbf{A Relaxed GAN Reconstruction Formulation.}
Despite the gap between the approximated manifold and the real one,
a well trained GAN generator still covers rich statistics of natural images.
In order to make use of these statistics while avoiding the aforementioned limitation,
we propose a relaxed GAN reconstruction formulation by 
allowing parameters $\vtheta$ of the generator to be moderately fine-tuned along with the latent vector $\vz$.
Such a relaxation on $\vtheta$ gives rise to an updated objective:
\begin{align}
\vtheta^*, \vz^* &= \argmin_{\vtheta, \vz} \cL(\hat{\vx}, \phi(G(\vz;\vtheta))), \quad \vx^* = G(\vz^*;\vtheta^*). \label{eq:dgp}
\end{align}
We refer to this updated objective as Deep Generative Prior (DGP).
With this relaxation, DGP significantly improves the chance of locating an optimal $\vx^*$ for $\hat{\vx}$,
as fitting the generator to a single image is much more achievable than fully capturing a data manifold.
Note that the generative prior buried in $G$, \eg,~its ability to output faithful natural images, might be deteriorated during the fine-tuning process.
The key to preserve the generative prior lies in the design of 
a good distance metric $\cL$ and a proper optimization strategy.

\begin{figure*}[t!]
	\centering
	\includegraphics[width=\linewidth]{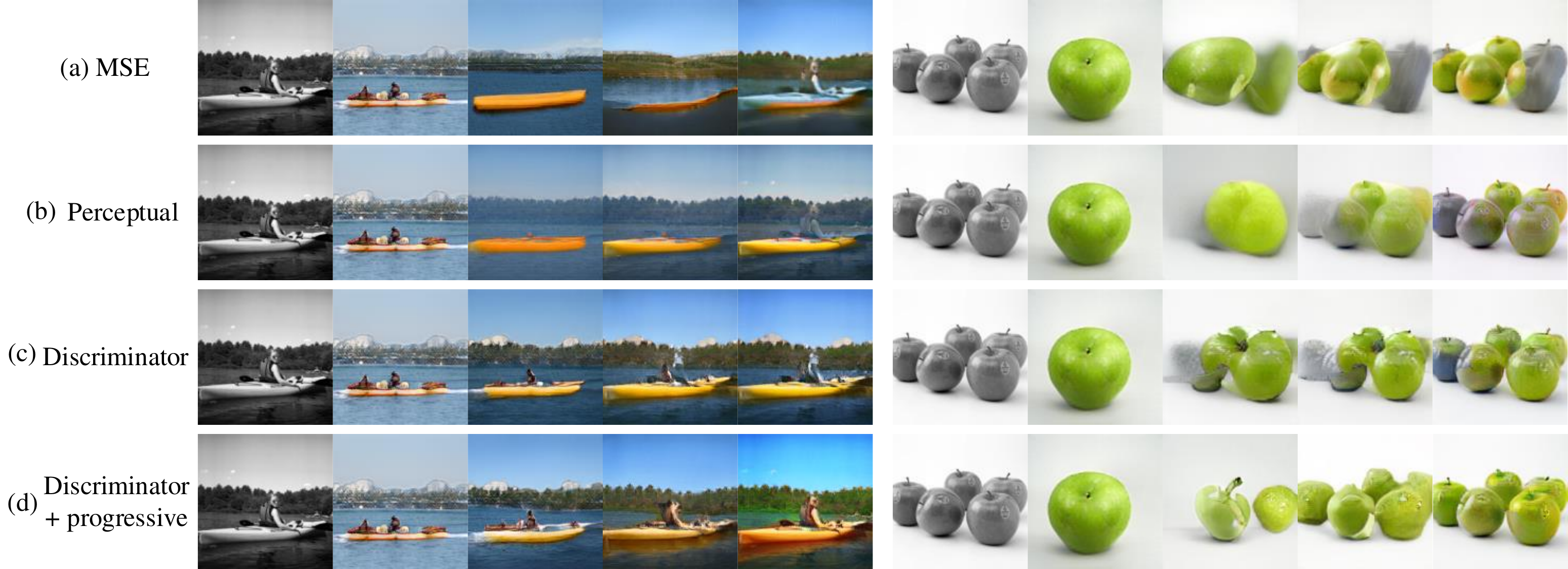}
	\vspace{-0.7cm}
	\caption{Comparison of different loss types when fine-tuning the generator to reconstruct the image}
	\vspace{-0.3cm}
	\label{loss_type}
\end{figure*}

\subsection{Discriminator Guided Progressive Reconstruction}
\label{sec:reconstruction}
To fit the GAN generator to the input image $\hat{\vx}$ while retaining a natural output,
in this section we introduce a discriminator based distance metric,
and a progressive fine-tuning strategy.

\vspace{3pt}
\noindent\textbf{Discriminator Matters.}
Given an input image $\hat{\vx}$, DGP will start with an initial latent vector $\vz_0$.
In practice, we obtain $\vz_0$ by randomly sampling a few hundreds of candidates from the latent space $\cZ$ 
and selecting the one that its corresponding image $G(\vz;\vtheta)$ best resembles $\hat{\vx}$ under the metric $\cL$ we used in Eq.\eqref{eq:dgp}.
As shown in Fig.~\ref{loss_type},
the choice of $\cL$ significantly affects the optimization of Eq.\eqref{eq:dgp}.
Existing literature often adopts the Mean-Squared-Error (MSE) \cite{ulyanov2018deep} 
or the AlexNet/VGGNet based Perceptual loss \cite{johnson2016perceptual,zhu2016generative} as $\cL$,
which respectively emphasize the pixel-wise appearance and the low-level/mid-level texture.
However, we empirically found using these metrics in Eq.\eqref{eq:dgp} often cause unfaithful outputs at the beginning of optimization, leading to sub-optimal results at the end.
We thus propose to replace them with a discriminator-based distance metric,
which measures the L1 distance in the \emph{discriminator feature space}:
\begin{align}
\cL(\vx_1, \vx_2) = \sum_{i \in \cI} \| D(\vx_1, i), D(\vx_2, i) \|_1, \label{eq:distance}
\end{align}
where $\vx_1$ and $\vx_2$ are two images, corresponding to $\hat{\vx}$ and $\phi(G(\vz;\vtheta))$ in Eq.\ref{eq:reconstruction} and Eq.\ref{eq:dgp}, and $D$ is the discriminator that is coupled with the generator.
$D(\vx, i)$ returns the feature of $\vx$ at $i$-block of \textit{D}, and $\cI$ is the index set of used blocks.
Compared to the AlexNet/VGGNet based perceptual loss,
the discriminator $D$ is trained along with $G$, instead of being trained for a separate task.
$D$, being a distance metric, thus is less likely to break the parameter structure of $G$, 
as they are well aligned during the pre-training.
Moreover, we found the optimization of DGP using such a distance metric visually works like an image morphing process.
\eg,~as shown in Fig.~\ref{loss_type}, the person on the boat is preserved and all intermediate outputs are all vivid natural images.
It is worth pointing out again while the feature matching loss is not new, this is the first time it serves as a regularizer during GAN reconstruction.

\begin{figure}[t]
    \begin{minipage}{0.59\textwidth} \hspace{0.03\textwidth}
    \includegraphics[width=6.2cm,height=3.9cm]{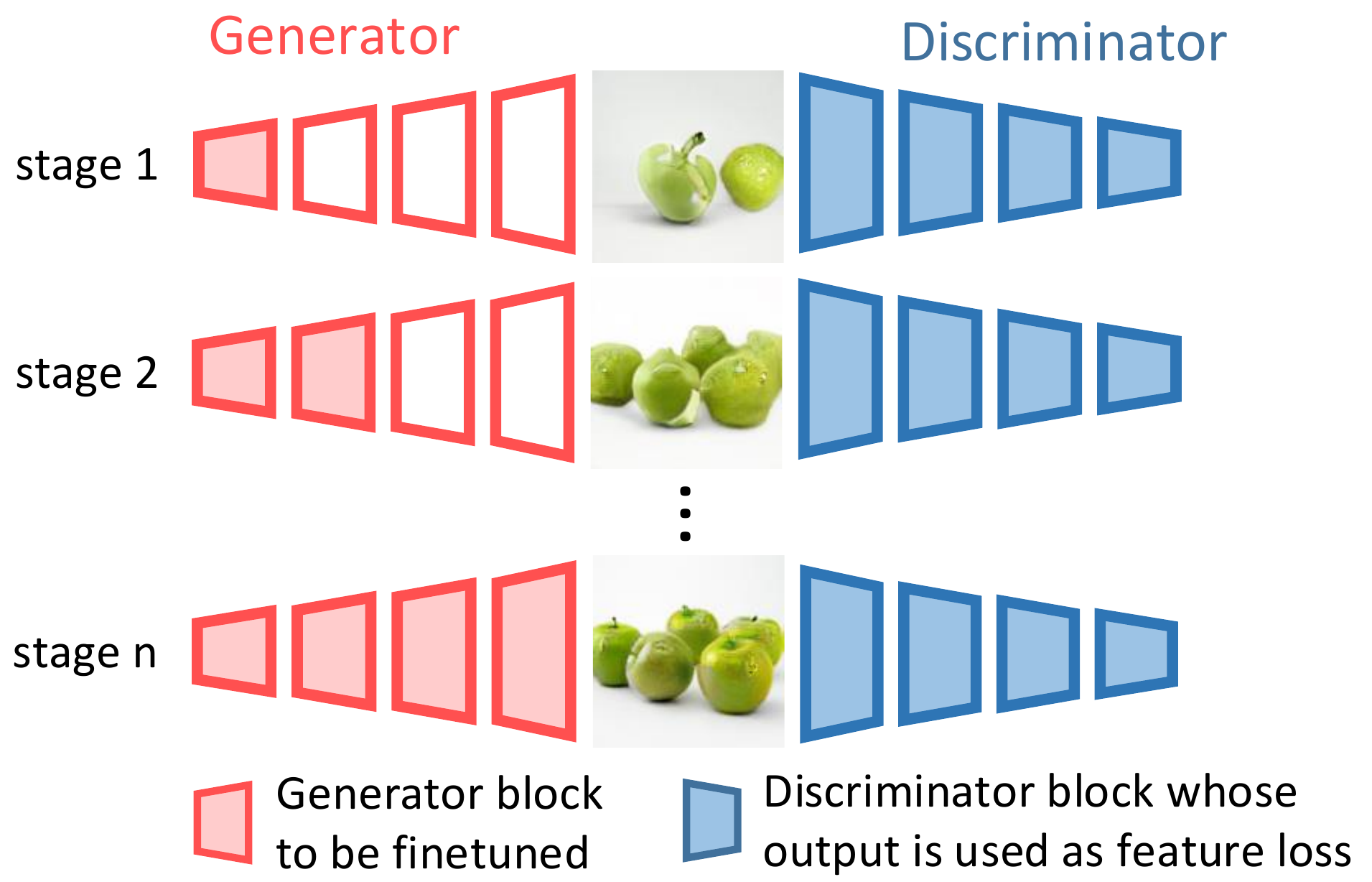}
    \end{minipage}\hspace{-0.04\textwidth}
    \begin{minipage}{0.39\textwidth} 
    \caption{\label{progressive} Progressive reconstruction of the generator can better preserves the consistency between missing and existing semantics in comparison to simultaneous fine-tuning on all the parameters at once. Here the list of images shown in the middle are the outputs of the generator in different fine-tuning stages. }
    \label{fig:example}
    \end{minipage}
\vspace{-0.4cm}
\end{figure}

\vspace{3pt}

\noindent\textbf{Progressive Reconstruction.}
Typically, we will fine-tune all parameters of $\vtheta$ simultaneously during the optimization of Eq.\eqref{eq:dgp}.
However, we observe an adverse effect of \emph{`information lingering'}, 
where missing semantics (\eg~color) do not shift along with existing context.
Taking Fig.~\ref{loss_type} (c) as an example, 
the leftmost apple fails to inherit the green color of the initial apple when it emerges.
One possible reason is deep blocks of the generator $G$ start to match low-level textures before high-level configurations are completely aligned.
To overcome this problem, we propose a progressive reconstruction strategy for some restoration tasks.

Specifically, as illustrated in Fig.~\ref{progressive},
we first fine-tune the shallowest block of the generator, and gradually continue with blocks at deeper depths,
so that DGP can control the global configuration at the beginning and gradually shift its attention to details at lower levels.
A demonstration of the proposed strategy is included in Fig.~\ref{loss_type} (d),
where DGP splits the apple from one to two at first,
then increases the number to five,
and finally refines the details of apples.
Compared to the non-progressive counterpart,
such a progressive strategy better preserves the consistency between missing and existing semantics.

%% file: sections/experiments.tex
\section{Applications}

We first compare our method with other GAN inversion methods for reconstruction, and then show the application of DGP in a number of image restoration and image manipulation tasks.
We adopt a BigGAN~\cite{brock2018large} to progressively reconstruct given images based on discriminator feature loss. 
The BigGAN is pre-trained on the ImageNet training set for conditional image synthesis.
BigGAN is selected due to its excellent performance in image generation. Other GANs are possible.
For dataset, we use the ImageNet~\cite{deng2009imagenet} validation set that has not been observed by BigGAN.
To quantitatively evaluate our method on image restoration tasks, we test on 1k images from the ImageNet validation set, where the first image for each class is collected to form the test set.
We recommend readers to refer to the Appendix for implementation details and more qualitative results.

\vspace{3pt}
\noindent\textbf{Comparison with other GAN-inversion methods.}
To begin with, we compare with other GAN-inversion methods~\cite{creswell2018inverting,Albright2019CVPRWorkshops,zhu2016generative,bau2019seeing} for image reconstruction.
As shown in Table~\ref{table_reconstruction}, our method achieves a very high PSNR and SSIM scores, outperforming other GAN-inversion methods by a large margin.
It can be seen from Fig.~\ref{fig:gan_inversion} that conventional GAN-inversion methods like ~\cite{zhu2016generative,bau2019seeing} suffer from obvious mismatches between reconstructed images and the target one, where the details or even contents are not well aligned.
In contrast, the reconstruction error of DGP is almost visually imperceptible.
More qualitative examples are provided in the Appendix.
%
%
In the following sections we show that our method also well exploits the generative prior in various applications.

\setlength{\tabcolsep}{4pt}
\begin{table}[!t]
	\begin{center}
		\caption{Comparison with other GAN-inversion methods, including (a) optimizing latent vector~\cite{creswell2018inverting,Albright2019CVPRWorkshops}, (b) learning an encoder~\cite{zhu2016generative}, (c) a combination of (a)(b)~\cite{zhu2016generative}, and (d) adding small perturbations to early stages based on (c)~\cite{bau2019seeing}. We reported PSNR, SSIM, and MSE of image reconstruction. The results are evaluated on the 1k ImageNet validation set}
		\vspace{-0.3cm}
		\resizebox{7cm}{0.9cm}{
	\begin{tabular}{c|ccccc}
		\Xhline{2\arrayrulewidth}
		& (a) & (b) & (c) & (d) & Ours \\ \hline\hline
		PSNR$\uparrow$ & 15.97 & 11.39    & 16.46        & 22.49 & \textbf{32.89} \\
		SSIM$\uparrow$ & 46.84 & 32.08    & 47.78        & 73.17 & \textbf{95.95} \\
		MSE$\downarrow$ ($\times$e-3) & 29.61 & 85.04    & 28.32  & 6.91 & \textbf{1.26} \\
		\Xhline{2\arrayrulewidth}
	\end{tabular}
	}
		\label{table_reconstruction}
	\end{center}
\vspace{-0.5cm}
\end{table}
\setlength{\tabcolsep}{1.4pt}

\begin{figure*}[t!]
	\centering
	\includegraphics[width=\linewidth]{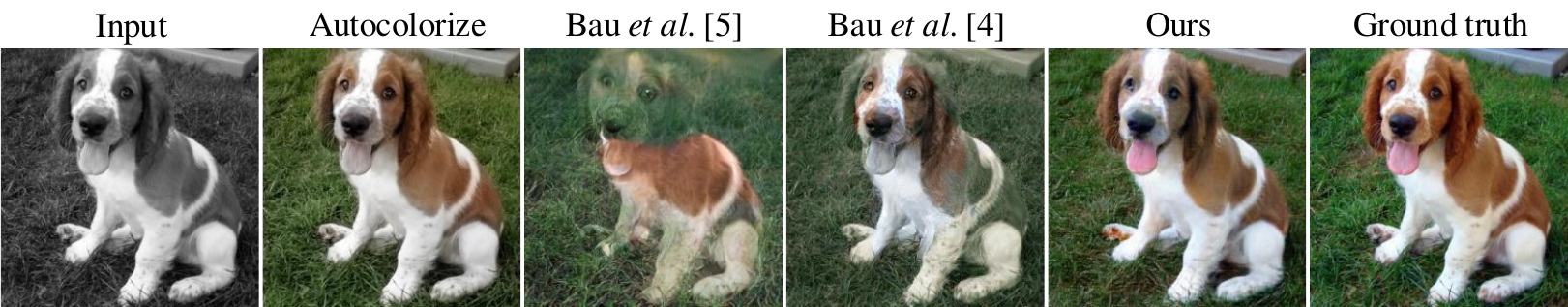}
	\vspace{-0.7cm}
	\caption{\textbf{Colorization}. Qualitative comparison of Autocolorize~\cite{larsson2016learning}, other GAN-inversion methods~\cite{bau2019seeing}\cite{bau2019semantic}, and our DGP}
	\vspace{-0.4cm}
	\label{colorization}
\end{figure*}

\vspace{-0.2cm}
\subsection{Image Restoration}
\vspace{-0.1cm}

\noindent\textbf{Colorization.}
Image colorization aims at restoring a gray-scale image $\hat{\vx} \in \mathbb{R}^{H \times W}$ to a colorful image with RGB channels $\vx \in \mathbb{R}^{3 \times H \times W}$.
To obtain $\hat{\vx}$ from the colorful image $\vx$, the degradation transform $\phi$ is a graying transform that only preserves the brightness of $\vx$.
By taking this degradation transform to Eq.\eqref{eq:dgp}, the goal becomes finding the colorful image $\vx^*$ whose gray-scale image is the same as $\hat{\vx}$.
We optimize Eq.\eqref{eq:dgp} using back-propagation and the progressive discriminator based reconstruction technique in Section \ref{sec:reconstruction}.
Fig.~\ref{loss_type}(d) shows the reconstruction process.
Note that the colorization task only requires to predict the ``ab" dimensions of the Lab color space.
Therefore, we transform $\vx^*$ to the Lab space, and adopt its ``ab" dimensions as well as the given brightness dimension $\hat{\vx}$ to produce the final colorful image.

Fig.~\ref{colorization} presents the qualitative comparisons with the Autocolorize~\cite{larsson2016learning} method.
Note that Autocolorize is directly optimized to predict color from gray-scale images while our method does not adopt such task-specific training.
Despite so, our method is visually better or comparable to Autocolorize.
To evaluate the colorization quality, we report the classification accuracy of a ResNet50~\cite{he2016deep} model on the colorized images. 
The ResNet50 accuracy for Autocolorize~\cite{larsson2016learning}, Bau \etal\cite{bau2019seeing}, Bau \etal\cite{bau2019semantic}, and ours are 51.5\%, 56.2\%, 56.0\%, and 62.8\% respectively, showing that DGP outperforms other baselines on this perceptual metric.

\begin{figure*}[t!]
	\centering
	\includegraphics[width=\linewidth]{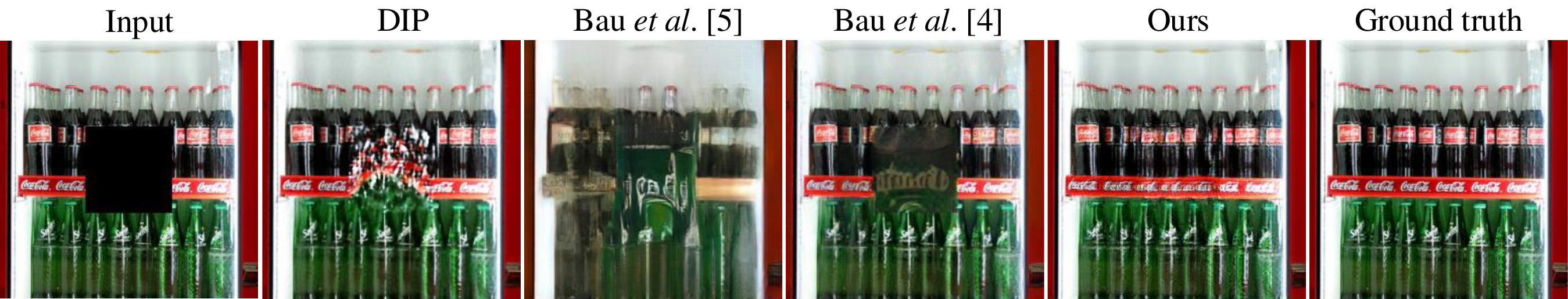}
	\vspace{-0.75cm}
	\caption{\textbf{Inpainting.} Compared with DIP and~\cite{bau2019seeing}\cite{bau2019semantic}, the proposed DGP could preserve the spatial coherence in image inpainting with large missing regions}
	\vspace{-0.3cm}
	\label{inpainting}
\end{figure*}

\begin{figure*}[t!]
	\centering
	\includegraphics[width=\linewidth]{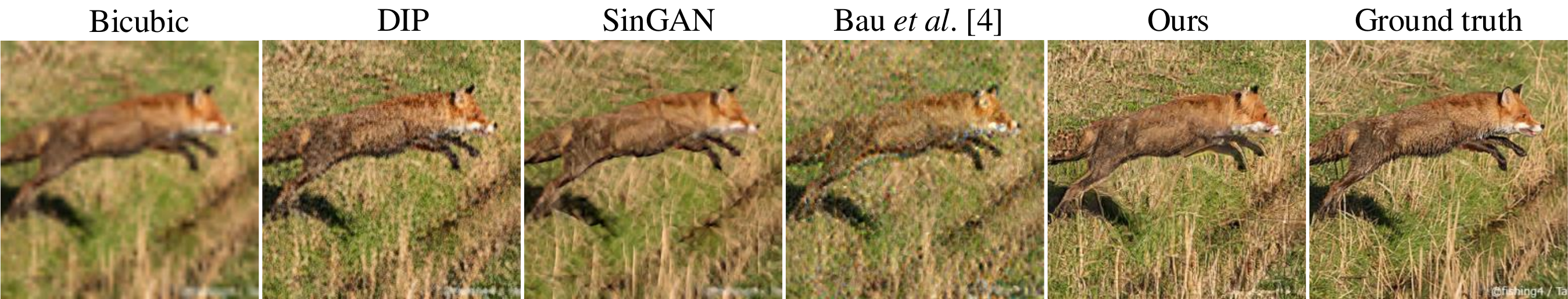}
	\vspace{-0.75cm}
	\caption{\textbf{Super-resolution ($\times 4$)} on $64 \times 64$ size images. 
		The comparisons of our method with DIP, SinGAN, and~\cite{bau2019semantic} are shown, where DGP produces sharper super-resolution results}
	\vspace{-0.1cm}
	\label{SR}
\end{figure*}

\vspace{3pt}
\noindent\textbf{Inpainting.}
The goal of image inpainting is to recover the missing pixels of an image.
The corresponding degradation transform is to multiply the original image with a binary mask $\vm$:
$\phi(\vx) = \vx \odot \vm$, where $\odot$ is Hadamard's product.
As before, we put this degradation transform to Eq.\eqref{eq:dgp}, and reconstruct target images with missing boxes.
Thanks to the generative image prior of the generator, the missing part tends to be recovered in harmony with the context, as illustrated in Fig.~\ref{inpainting}.
In contrast, the absence of a learned image prior would result in messy inpainting results, as in DIP.
Quantitative results indicate that DGP outperforms DIP and other GAN-inversion methods by a large margin, as Table \ref{table_inpainting} shows.

\setlength{\tabcolsep}{4pt}
\begin{table}[!t]
	\begin{center}
		\caption{Inpainting evaluation. We reported PSNR and SSIM of the inpainted area. The results are evaluated on the 1k ImageNet validation set}
		\vspace{-0.3cm}
		\begin{tabular}{c|ccccc}
			\Xhline{2\arrayrulewidth}
				   & DIP & Zhu \etal\cite{zhu2016generative} & Bau \etal\cite{bau2019seeing} & Bau \etal\cite{bau2019semantic} & Ours  \\ \hline\hline
			PSNR$\uparrow$ & 14.58 & 13.70 & 15.01 & 14.33 & \textbf{16.97} \\
			SSIM$\uparrow$  & 29.37 & 33.09 & 33.95 & 30.60 & \textbf{45.89} \\ \Xhline{2\arrayrulewidth}
		\end{tabular}
		\label{table_inpainting}
	\end{center}
\vspace{-0.5cm}
\end{table}
\setlength{\tabcolsep}{1.4pt}

\setlength{\tabcolsep}{4pt}
\begin{table}[!t]
	\begin{center}
		\caption{Super-resolution ($\times 4$) evaluation. We reported widely used NIQE~\cite{mittal2012making}, PSNR, and RMSE scores. The results are evaluated on the 1k ImageNet validation set. (MSE) and (D) indicate which kind of loss DGP is biased to use}
		\vspace{-0.2cm}
		\resizebox{8.6cm}{0.9cm}{
			\begin{tabular}{c|ccccccc}
				\Xhline{2\arrayrulewidth}
				& DIP  & SinGAN & Bau \etal\cite{bau2019semantic} & Ours (MSE) & Ours (D) \\ \hline\hline
				NIQE$\downarrow$  & 6.03   & 6.28 & 5.05 &  5.30  &  \textbf{4.90} \\
				PSNR$\uparrow$  & 23.02  & 20.80  & 19.89 & \textbf{23.30}   & 22.00    \\
				RMSE$\downarrow$  & 17.84  & 19.78 & 25.42 & \textbf{17.40}  & 20.09  \\
				 \Xhline{2\arrayrulewidth}
			\end{tabular}
		}
	\label{table:SR}
	\end{center}
\vspace{-0.7cm}
\end{table}
\setlength{\tabcolsep}{1.4pt}

\vspace{3pt}
\noindent\textbf{Super-Resolution.}
In this task, one is given with a low-resolution image $\hat{\vx} \in \mathbb{R}^{3 \times H \times W}$, and the purpose is to generate the corresponding high-resolution image $\vx \in \mathbb{R}^{3 \times fH \times fW}$, where $f$ is the upsampling factor.
In this case, the degradation transform $\phi$ is to downsample the input image by a factor $f$.
Following DIP~\cite{ulyanov2018deep}, we adopt the Lanczos downsampling operator in this work.

Fig.~\ref{SR} and Table \ref{table:SR} show the comparison of DGP with DIP, SinGAN, and Bau \etal\cite{bau2019semantic}.
Our method achieves sharper and more faithful super-resolution results than its counterparts.
For quantitative results, we could trade off between perceptual quality like NIQE and commonly used PSNR score by using different combination ratios of discriminator loss and MSE loss at the final fine-tuning stage.
For instance, when using higher MSE loss, DGP has excellent PSNR and RMSE performance, and outperforms other counterparts in all the metrics involved.
And the perceptual quality NIQE could be further improved by biasing towards discriminator loss.

\begin{figure}[t!]
	\centering
	\includegraphics[width=\linewidth]{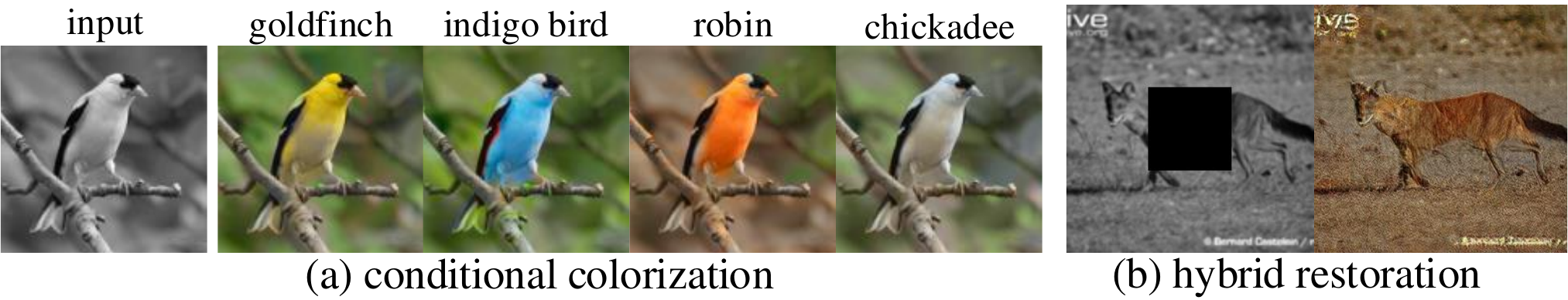}
	\vspace{-0.75cm}
	\caption{(a) Colorizing an image under different class conditions. (b) Simultaneously conduct colorization, inpainting, and super-resolution ($\times 2$)}
	\vspace{-0.1cm}
	\label{fig:conditional_hybrid}
\end{figure}

\begin{figure}[t!]
	\centering
	\includegraphics[width=\linewidth]{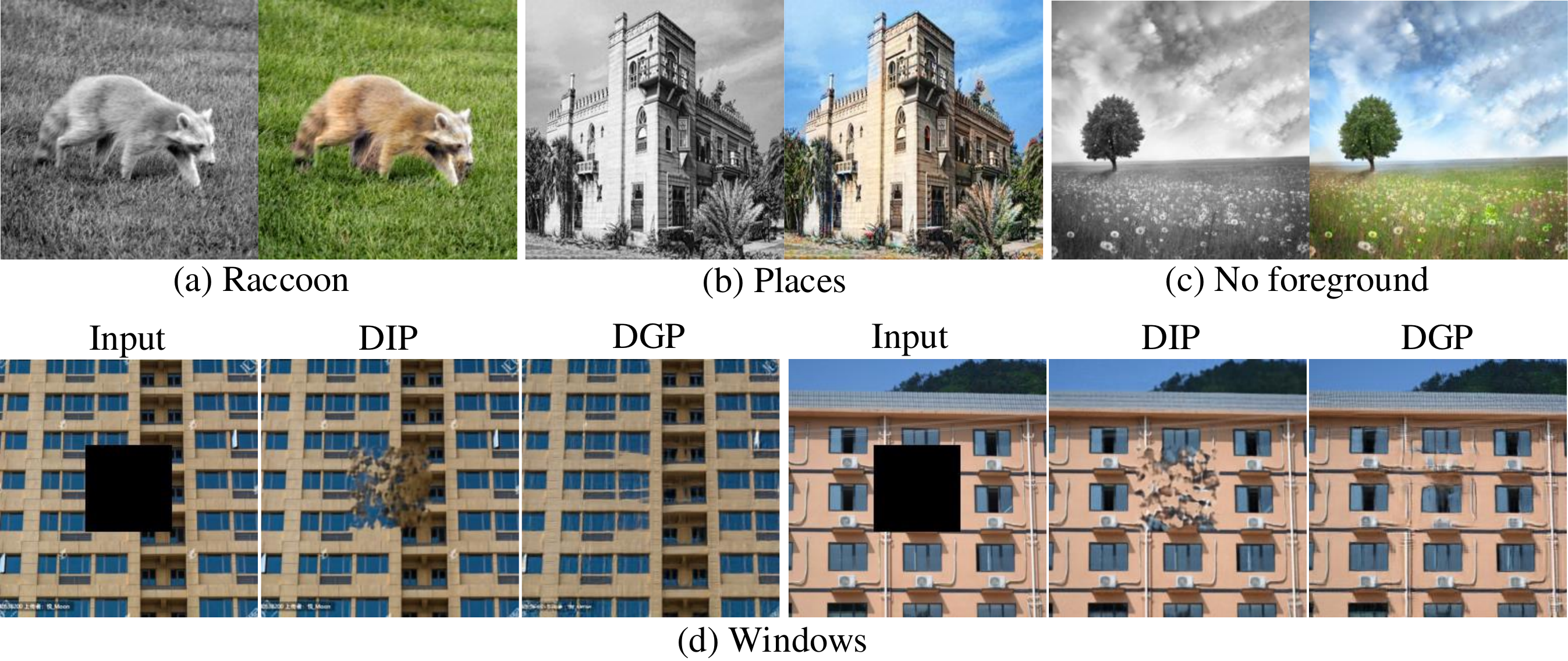}
	\vspace{-0.75cm}
	\caption{Evaluation of DGP on non-ImageNet images, including (a) `Raccoon', a category not belonging to ImageNet categories, (b) image from Places dataset~\cite{zhou2017places}, (c) image without foreground object, and (d) windows. (a)(c)(d) are scratched from Internet}
	\vspace{-0.4cm}
	\label{fig:generalization}
\end{figure}

\vspace{3pt}
\noindent\textbf{Flexibility of DGP.}
The generic paradigm of DGP provides more flexibility in restoration tasks.
For example, an image of gray-scale bird may have many possibilities when restored in the color space.
Since the BigGAN used in our method is a conditional GAN, we could achieve diversity in colorization by using different class conditions when restoring the image, as Fig.~\ref{fig:conditional_hybrid} (a) shows.
Furthermore, our method allows hybrid restoration, \ie, jointly conducting colorization, inpainting, and super-resolution.
This could naturally be achieved by using a composite of degrade transform $\phi(\vx) = \phi_a(\phi_b(\phi_c(\vx)))$, as shown in Fig.~\ref{fig:conditional_hybrid} (b).

\vspace{3pt}
\noindent\textbf{Generalization of DGP.}
We also test our method on images not belonging to ImageNet.
As Fig.\ref{fig:generalization} shows, DGP restores the color and missed patches of these images reasonably well.
Particularly, compared with DIP, DGP fills the missed patches to be well aligned with the context.
This indicates that DGP does capture the \textit{`spatial coherence'} prior of natural images, instead of memorizing the ImageNet dataset.
We scratch a small dataset with 18 images of windows, stones, and libraries to test our method, where DGP achieves 15.34 for PSNR and 41.53 for SSIM, while DIP has only 12.60 for PSNR and 21.12 for SSIM.

\setlength{\tabcolsep}{4pt}
\begin{table}[!t]
	\begin{center}
		\caption{Comparison of different loss type and fine-tuning strategy}
		\vspace{-0.3cm}
		\resizebox{9cm}{1.0cm}{
		\begin{tabular}{c|c|cccc}
			\Xhline{2\arrayrulewidth}
			Task                & Metric & MSE   & Perceptual   & Discriminator     & \makecell{Discriminator\\+Progressive}  \\ \hline\hline
			Colorization        & ResNet50$\uparrow$ & 49.1  &  53.9   &  56.8   &  \textbf{62.8}  \\ \hline
			\multirow{2}{*}{SR} & NIQE$\downarrow$   & 6.54  & 6.27  & 6.06  & \textbf{4.90}  \\
			& PSNR$\uparrow$   & 21.24 & 20.30 & 21.58 & \textbf{22.00} \\ \Xhline{2\arrayrulewidth}
		\end{tabular}
	}
		\label{table:ablation}
	\vspace{-0.6cm}
	\end{center}
\end{table}
\setlength{\tabcolsep}{1.4pt}

\vspace{3pt}
\noindent\textbf{Ablation Study.}
To validate the effectiveness of the proposed discriminator guided progressive reconstruction method, we compare different fine-tuning strategies in Table~\ref{table:ablation}.
There is a clear improvement of discriminator feature matching loss over MSE and perceptual loss, and the combination of the progressive reconstruction further boosts the performance.
Fig.~\ref{fig:gan_inversion}, Fig.~\ref{loss_type}, and Appendix provide qualitative comparisons. The results show that the progressive strategy effectively eliminates the `information lingering' artifacts.

\begin{figure}[t!]
	\centering
	\includegraphics[width=7.0cm]{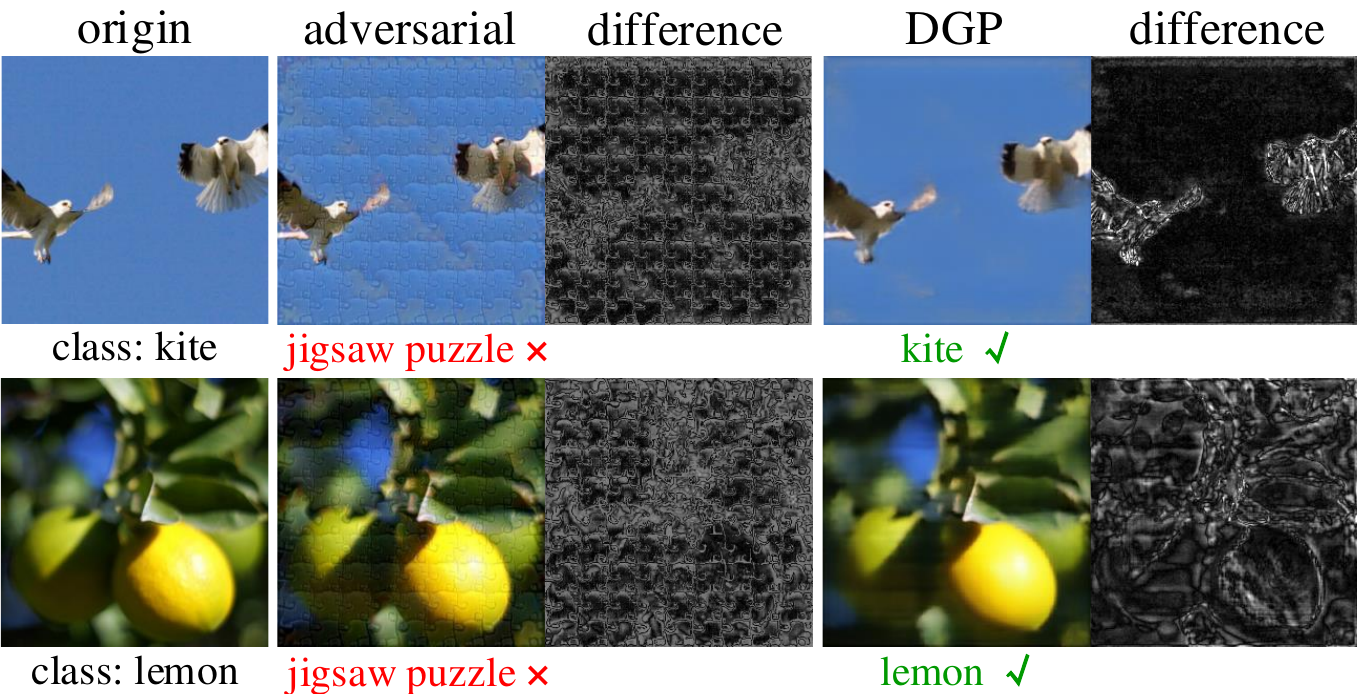}
	\vspace{-0.3cm}
	\caption{\textbf{Adversarial defense.} DGP is capable of filtering out unnatural perturbations in the adversarial samples by reconstructing them}
	\vspace{-0.1cm}
	\label{fig:defense}
\end{figure}

\setlength{\tabcolsep}{4pt}
\begin{table}[!t]
	\begin{center}
		\caption{Adversarial defense evaluation. We reported the classification accuracy of a ResNet50. The results are evaluated on the 1k ImageNet validation set}
		\vspace{-0.25cm}
		\resizebox{8.7cm}{0.7cm}{
			\begin{tabular}{l|cc|ccc}
				\Xhline{2\arrayrulewidth}
				method    & clean image & adversarial & DefenceGAN & DIP  & Ours  \\ \hline\hline
				top1 acc. (\%) & 74.9        & 1.4         &  0.2    & 37.5 & 41.3 \\
				top5 acc. (\%) & 92.7        & 12.0        &  1.4    & 61.2 & 65.9 \\ \Xhline{2\arrayrulewidth}
			\end{tabular}
		}
		\label{defense}
		\vspace{-0.7cm}
	\end{center}
\end{table}
\setlength{\tabcolsep}{1.4pt}

\vspace{3pt}
\noindent\textbf{Adversarial Defense.}
Adversarial attack methods aim at fooling a CNN classifier by adding a certain perturbation $\Delta \vx$ to a target image $\vx$~\cite{nguyen2015deep}.
In contrast, adversarial defense aims at preventing the model from being fooled by attackers.
Specifically, the work of DefenseGAN~\cite{samangouei2018defense} proposed to restore a perturbed image to a natural image by reconstructing it with a GAN.
It works well for simple data like MNIST, but would fail for complex data like ImageNet due to poor reconstruction.
Here we show the potential of DGP in adversarial defense under a black-box attack setting~\cite{baluja2017adversarial}, where the attacker does not have access to the classifier and defender.

For adversarial attack, the degradation transform is $\phi(\vx) = \vx + \Delta \vx$, where $\Delta \vx$ is the perturbation generated by the attacker.
Since calculating $\phi(\vx)$ is generally not differentiable, here we adopt DGP to directly reconstruct the adversarial image $\hat{\vx}$.
To prevent $\vx^*$ from overfitting to $\hat{\vx}$, we stop the reconstruction when the MSE loss reaches 5e-3.
We adopt the adversarial transformation networks attacker~\cite{baluja2017adversarial} to produce the adversarial samples\footnote[1]{We use the code at \textit{https://github.com/pfnet-research/nips17-adversarial-attack}}.

As Fig.~\ref{fig:defense} shows, the generated adversarial image contains unnatural perturbations, leading to misclassification for a ResNet50~\cite{he2016deep}.
After reconstructing the adversarial samples using DGP, the perturbations are largely alleviated, and the samples are thus correctly classified.
The comparisons of our method with DefenseGAN and DIP are shown in Table \ref{defense}.
DefenseGAN yields poor defense performance due to inaccurate reconstruction.
And DGP outperforms DIP, thanks to the learned image prior that produces more natural restored images.

\vspace{-0.3cm}

\subsection{Image Manipulation}

\begin{figure}[t!]
	\centering
	\includegraphics[width=8.8cm]{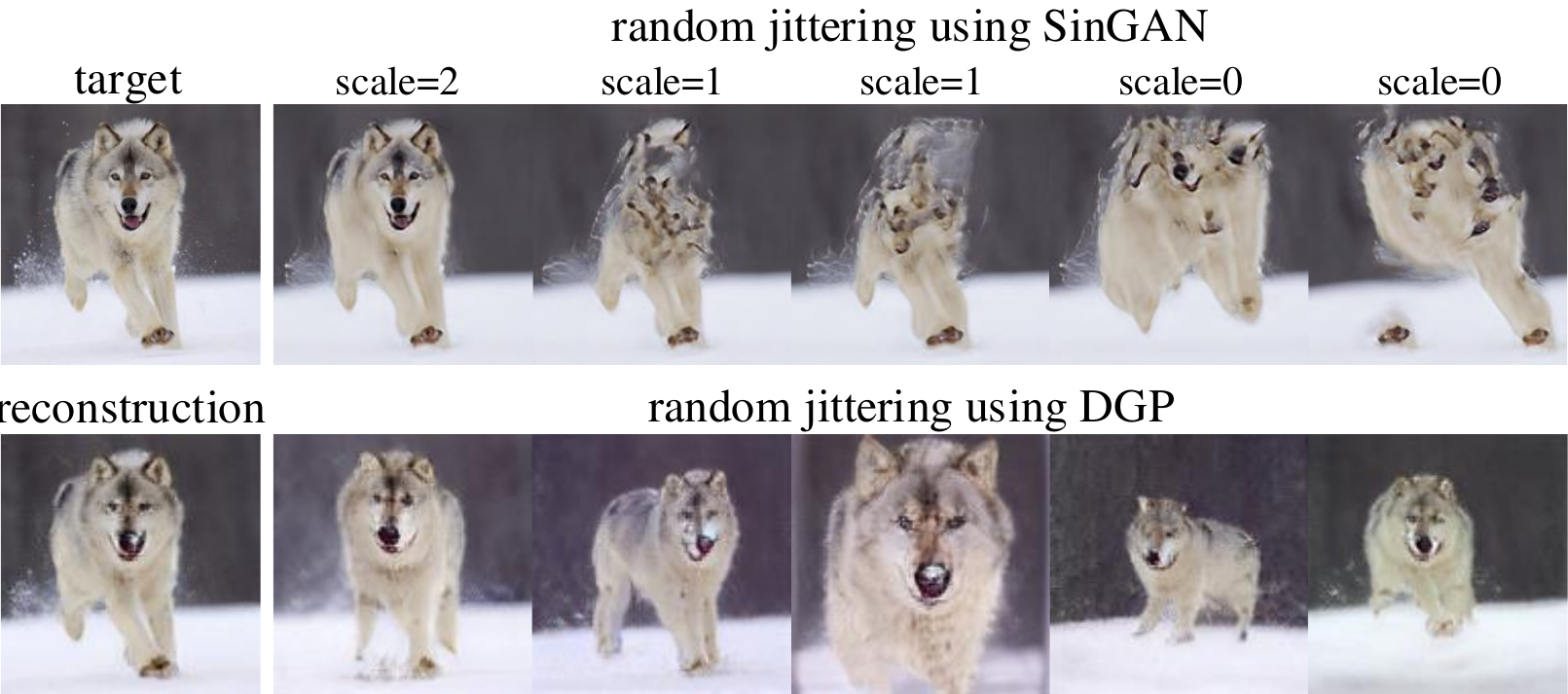}
	\vspace{-0.3cm}
	\caption{Comparison of \textbf{random jittering} using SinGAN (above) and DGP (below)}
	\vspace{-0.3cm}
	\label{fig:jitter}
\end{figure}

Since DGP enables precise GAN reconstruction while preserving the generative property, 
it becomes straightforward to apply the fascinating capabilities of GAN to real images like random jittering, image morphing, and category transfer.
In this section, we show the application of our method in these image manipulation tasks.

\vspace{3pt}
\noindent\textbf{Random Jittering.}
We show the random jittering effects of DGP, and compare it with SinGAN.
Specifically, after reconstructing a target image using DGP, we add Gaussian noise to the latent vector $\vz^*$ and see how the output changes.
As shown in Fig.~\ref{fig:jitter}, the dog in the image changes in pose, action, and size, where each variant looks like a natural shift of the original image.
For SinGAN, however, the jittering effects seem to preserve some texture, but losing the concept of `dog'.
This is because it cannot learn a valid representation of dog by looking at only one dog.
In contrast, in DGP the generator is fine-tuned in a moderate way such that the structure of image manifold captured by the generator is well preserved.
Therefore, perturbing $\vz^*$ corresponds to shifting the image in the natural image manifold.

\vspace{3pt}
\noindent\textbf{Image Morphing.}
The purpose of image morphing is to achieve a visually sound transition from one image to another.
Given a GAN generator \textit{G} and two latent vectors $\vz_A$ and $\vz_B$, morphing between $G(\vz_A)$ and $G(\vz_B)$ could naturally be done by interpolating between $\vz_A$ and $\vz_B$.
In the case of DGP, however, reconstructing two target images $\vx_A$ and $\vx_B$ would result in two generators $G_{\theta_A}$ and $G_{\theta_B}$, and the corresponding latent vectors $\vz_A$ and $\vz_B$.
Inspired by \cite{wang2019deep}, to morph between $\vx_A$ and $\vx_B$, we apply linear interpolation to both the latent vectors and the generator parameters:
$\vz = \lambda \vz_A + (1 - \lambda) \vz_B, \vtheta = \lambda \vtheta_A + (1 - \lambda) \vtheta_B, \lambda \in (0, 1)$,
and generate images with the new $\vz$ and $\vtheta$.

\begin{figure*}[t!]
	\centering
	\includegraphics[width=\linewidth]{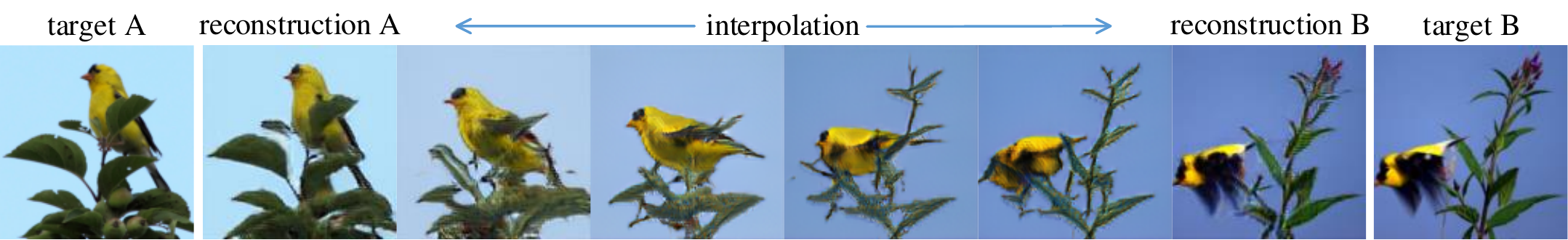}
	\vspace{-0.7cm}
	\caption{\textbf{Image morphing.} Our method achieves visually realistic image morphing effects} 
	\vspace{-0.25cm}
	\label{morphing}
\end{figure*}


As Fig.~\ref{morphing} shows, our method enables highly photo-realistic image morphing effects.
Despite the existence of complex backgrounds, the imagery contents shift in a natural way.
To quantitatively evaluate image morphing quality, we apply image morphing to every consecutive image pairs for each class in the ImageNet validation set, and collect the intermediate images where $\lambda = 0.5$.
For 50k images with 1k classes, this would create 49k generated images.
We evaluate the image quality using Inception Score (IS)~\cite{salimans2016improved}, and compare DGP with DIP, which adopts a similar network interpolation strategy.
Finally, DGP achieves a satisfactory IS, 59.9, while DIP fails to create valid morphing results, leading to only 3.1 of IS.


\begin{figure}[t!]
	\centering
	\includegraphics[width=10cm]{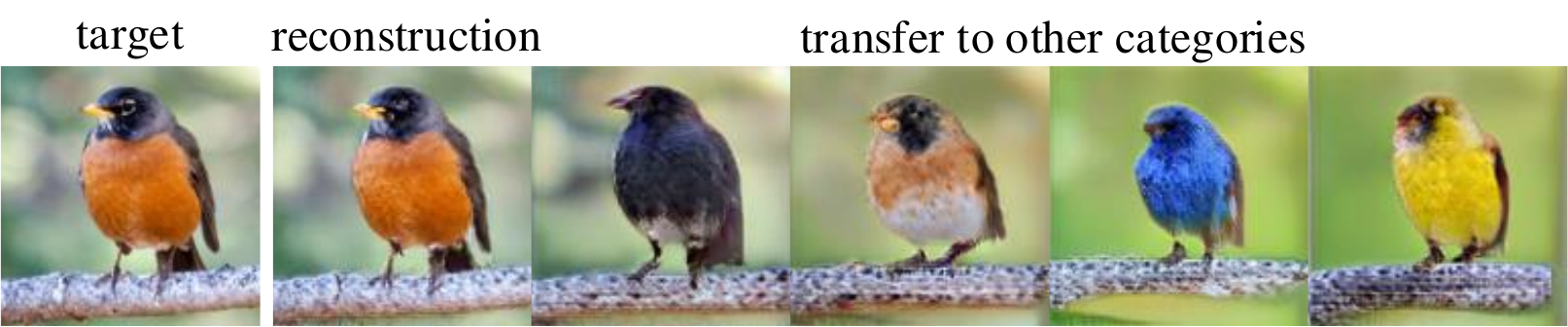}
	\vspace{-0.3cm}
	\caption{\textbf{Category transfer.} DGP enables the editing of semantics of objects in images}
	\vspace{-0.4cm}
	\label{category}
\end{figure}

\vspace{3pt}
\noindent\textbf{Category Transfer.}
In conditional GAN, the class condition controls the content to be generated.
So after reconstructing a given image via DGP, we can manipulate its content by tweaking the class condition.
Fig.~\ref{fig:intro} (f) and Fig.~\ref{category} present examples of transferring the object category of given images.
Our method can transfer the dog and bird to various other categories without changing the pose, size, and image configurations.



%% file: sections/conclusion.tex
\section{Conclusion}

To summarise, we have shown that a GAN generator trained on massive natural images could be used as a generic image prior, namely deep generative prior (DGP).
Embedded with rich knowledge on natural images, DGP could be used to restore the missing information of a degraded image by progressively reconstructing it under the discriminator metric.
Meanwhile, such reconstruction strategy addresses the challenge of GAN-inversion, achieving multiple visually realistic image manipulation effects.
Our results uncover the potential of a universal image prior captured by a GAN in image restoration and manipulation.


\vspace{7pt}
\noindent\textbf{Acknowledgment}
We would like to thank Xintao Wang for helpful discussions.

%% file: sections/appendix.tex


\setcounter{section}{0}
\renewcommand\thesection{\Alph{section}}

\section*{\fontsize{15}{15}\selectfont Appendix}

In this appendix, we provide more qualitative results and the implementation details in our experiments.
Readers can see restoration and manipulation videos at our \href{https://github.com/XingangPan/deep-generative-prior}{github repo}.

\section{Qualitative Examples}

\begin{figure}[h]
	\centering
	\includegraphics[width=\linewidth]{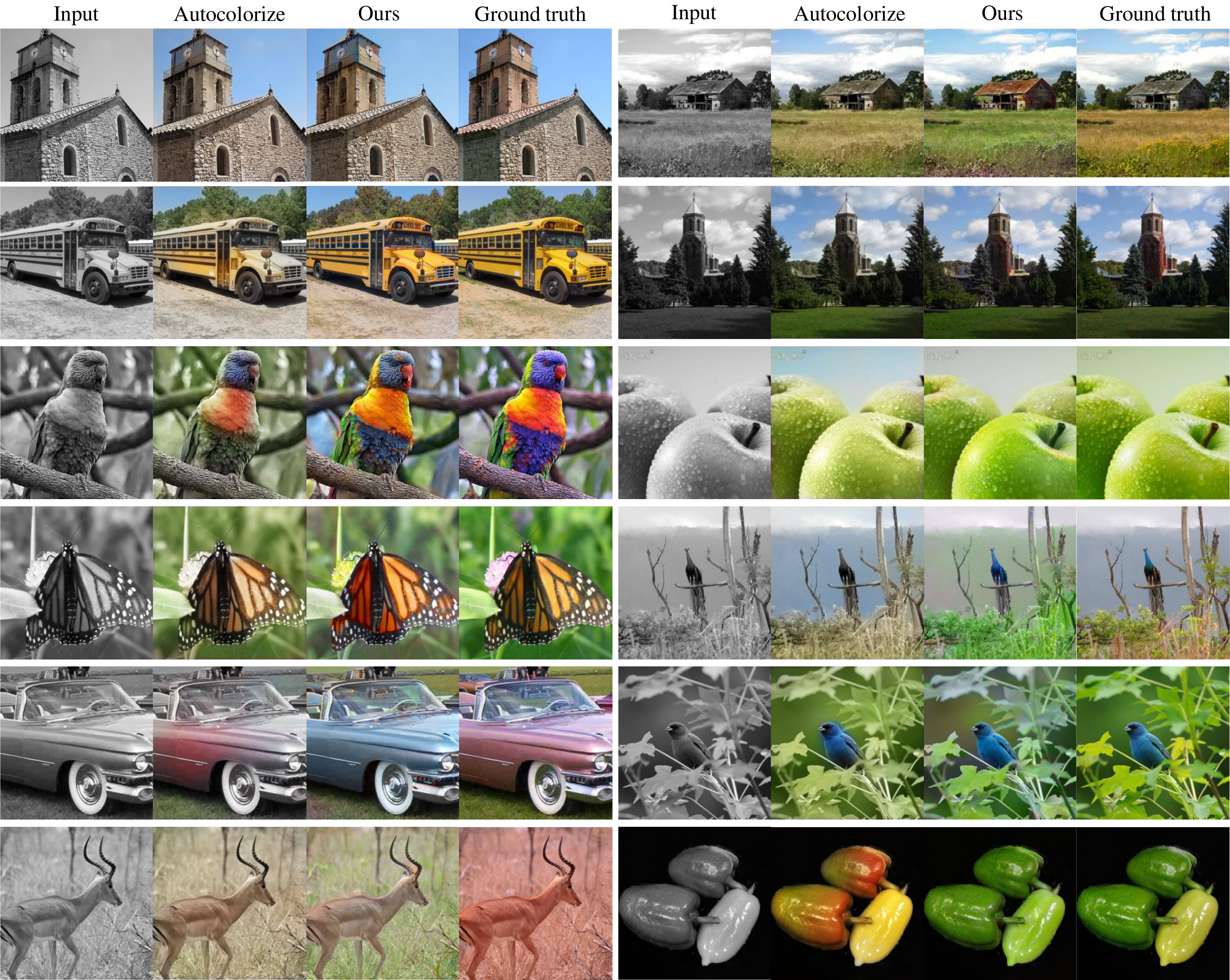}
	\caption{\textbf{Colorization.} This is an extension of Fig.5 in the main paper.}
	\label{sup:colorization}
\end{figure}

\begin{figure}[h!]
	\centering
	\vspace{-0.4cm}
	\includegraphics[width=\linewidth]{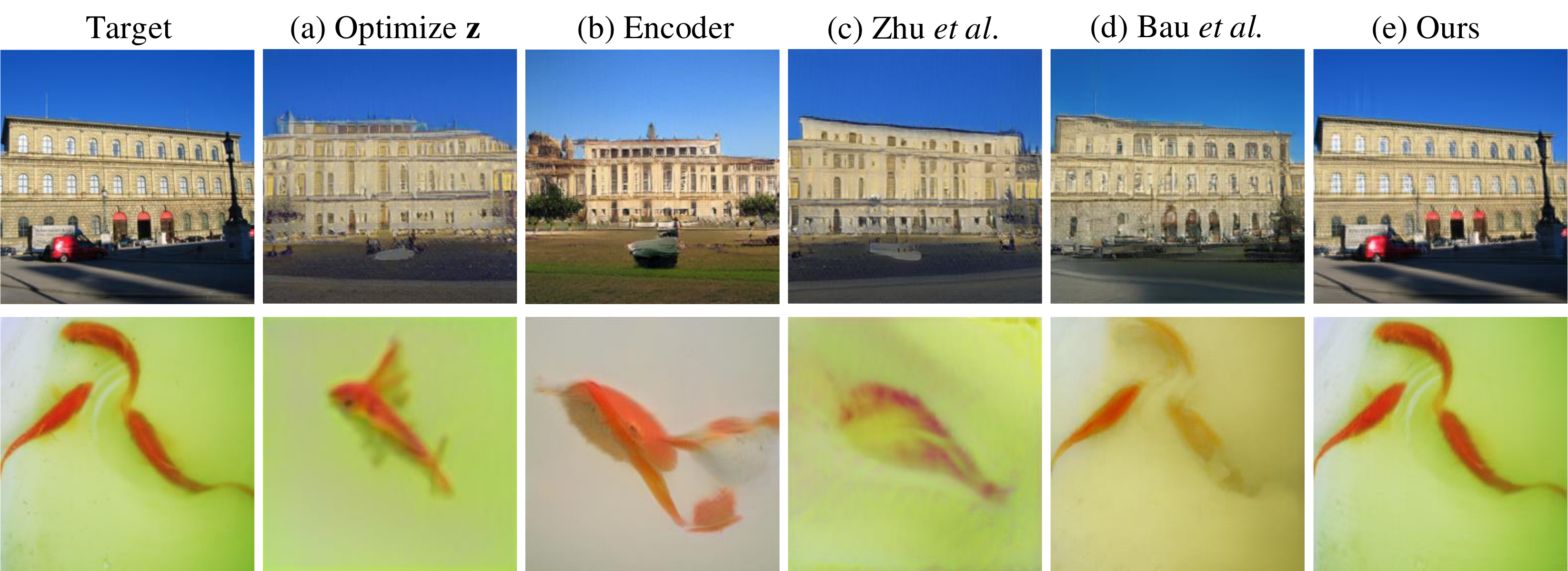}
	\vspace{-0.6cm}
	\caption{\textbf{Image reconstruction}. We compare our method with other GAN-inversion methods including (a) optimizing latent vector~\cite{creswell2018inverting,Albright2019CVPRWorkshops}, (b) learning an encoder~\cite{zhu2016generative}, (c) a combination of (a)(b)~\cite{zhu2016generative}, and (d) adding small perturbations to early stages based on (c)~\cite{bau2019seeing}. }
	\label{sup:reconstruct}
	\vspace{-0.5cm}
\end{figure}

\begin{figure}[h!]
	\centering
	\vspace{-0.4cm}
	\includegraphics[width=\linewidth]{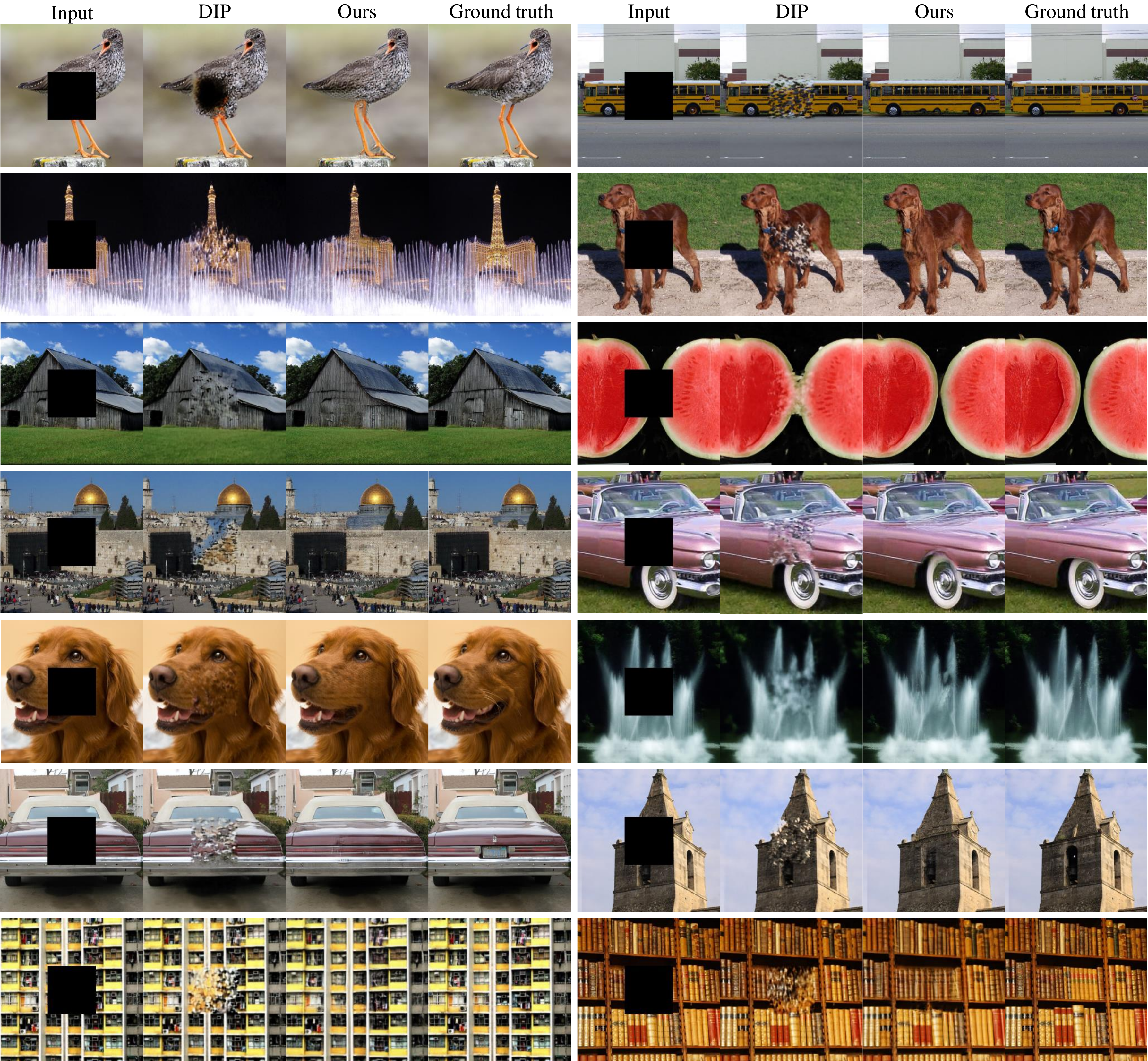}
	\vspace{-0.6cm}
	\caption{\textbf{Inpainting.} This is an extension of Fig.6 in the main paper. The proposed DGP tends to recover the missing part in harmony with the context. Images of the last row are scratched from the Internet. }
	\label{sup:inpainting}
	\vspace{-0.5cm}
\end{figure}

\begin{figure*}[h!]
	\centering
	\vspace{-0.4cm}
	\includegraphics[width=\linewidth]{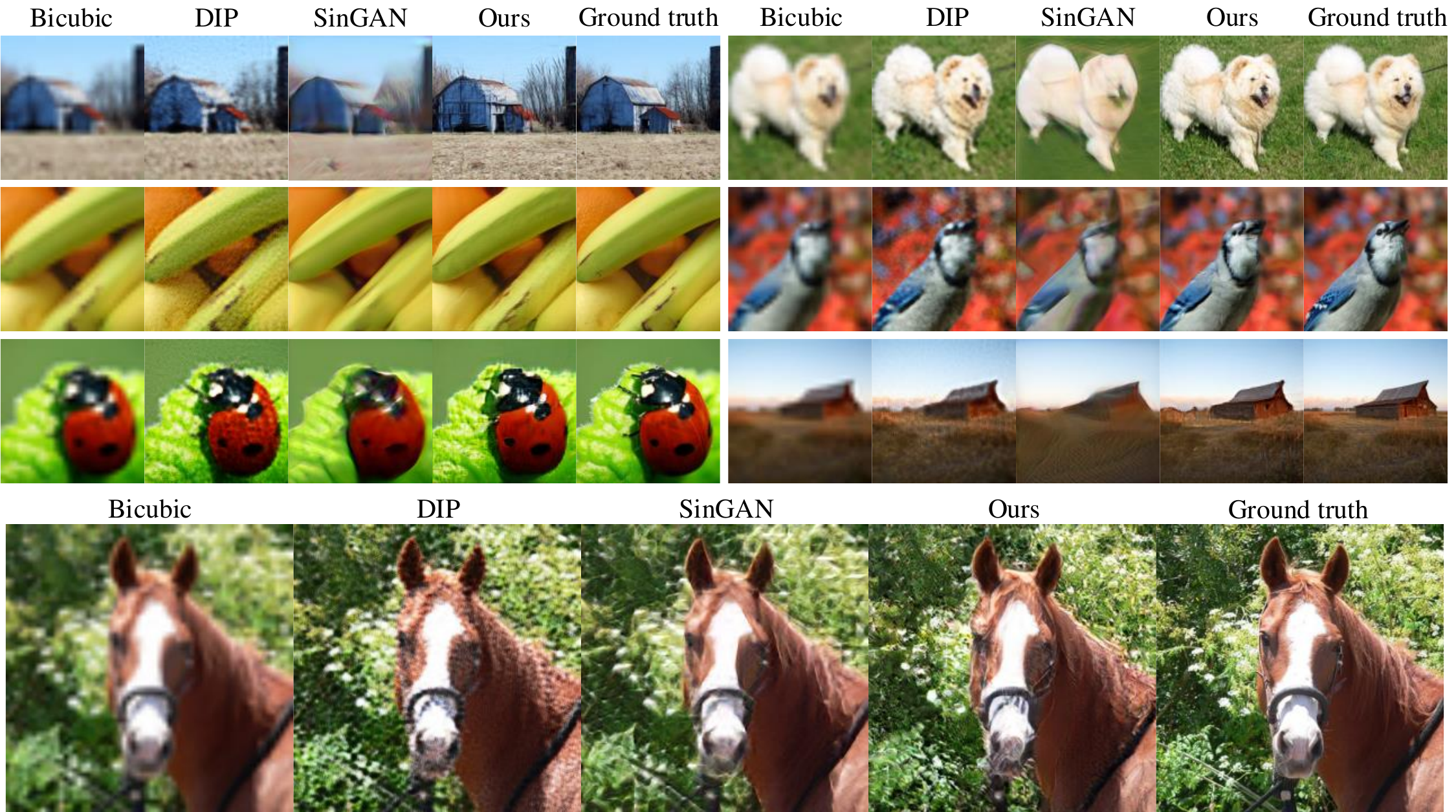}
	\vspace{-0.6cm}
	\caption{\textbf{Super-resolution ($\times 4$)} on $32 \times 32$ (above) and $64 \times 64$ (below) size images. This is an extension of Fig.7 in the main paper. }
	\label{sup:SR}
	\vspace{-0.5cm}
\end{figure*}

\begin{figure*}[h!]
	\centering
	\vspace{-0.4cm}
	\includegraphics[width=11cm]{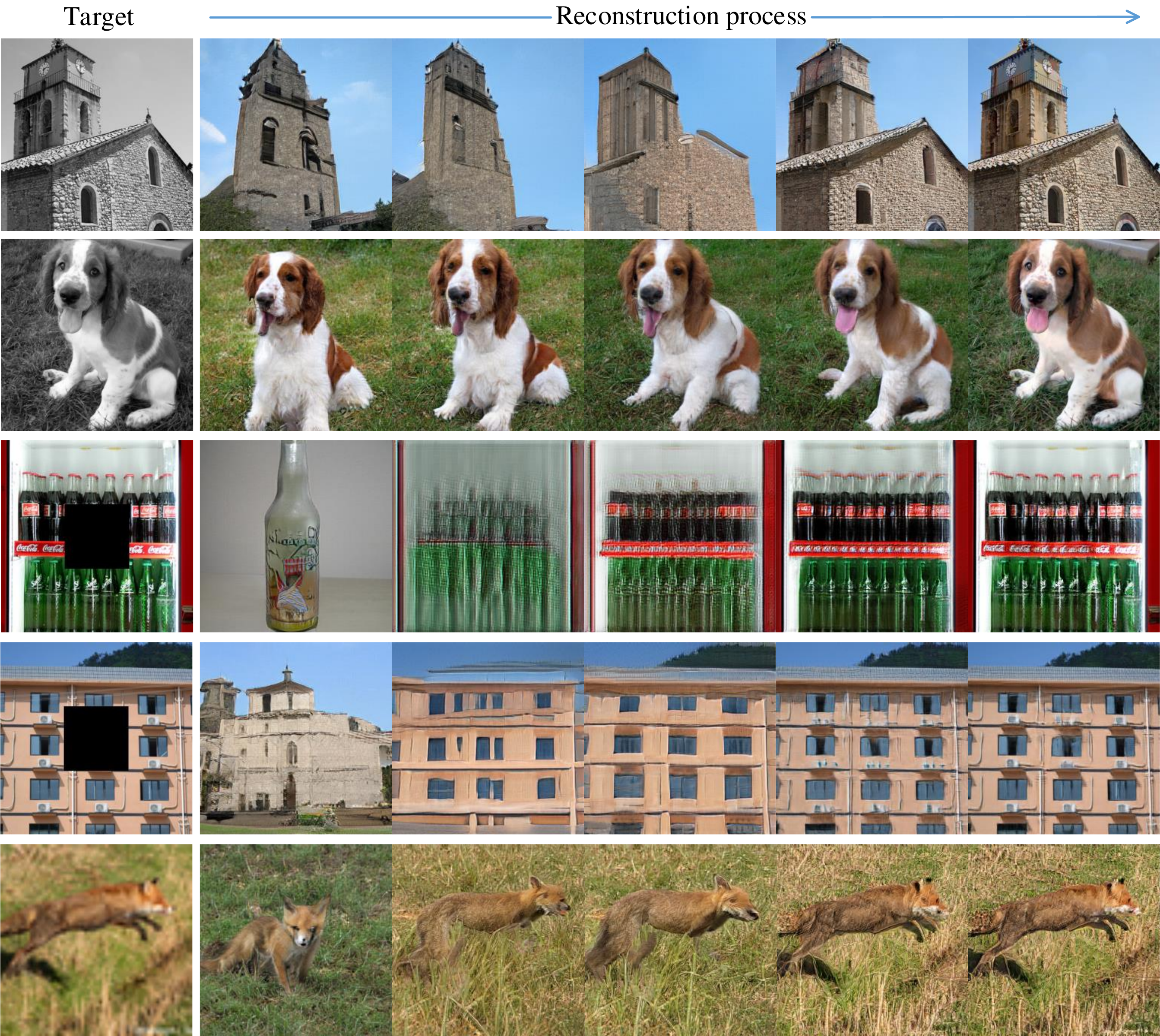}
	\vspace{-0.2cm}
	\caption{The reconstruction process of DGP in various image restoration tasks. }
	\label{sup:process}
	\vspace{-0.5cm}
\end{figure*}

\begin{figure}[h!]
	\centering
	\vspace{-0.4cm}
	\includegraphics[width=10cm]{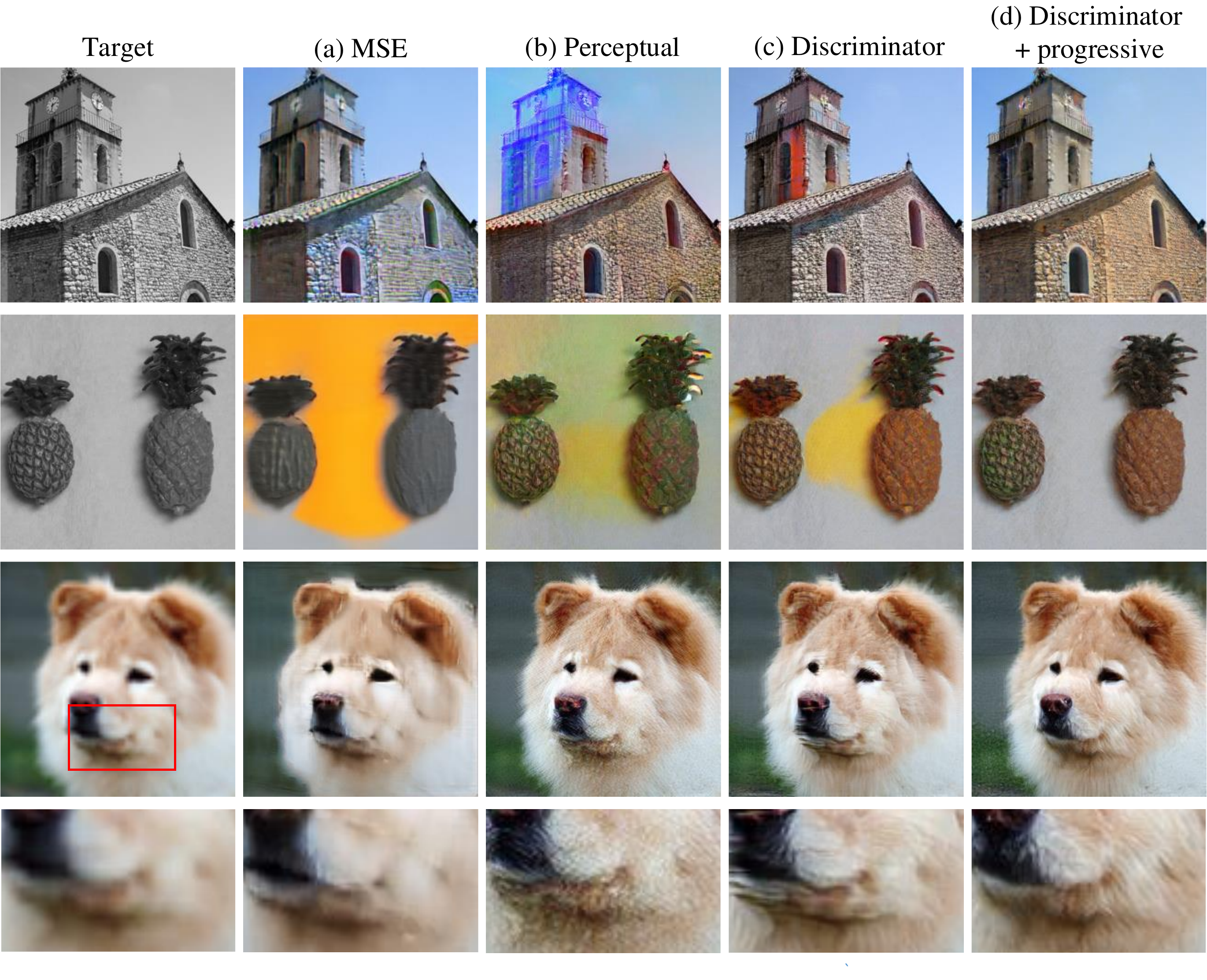}
	\vspace{-0.5cm}
	\caption{Comparison of different loss types and optimization techniques in colorization and super-resolution, including (a) MSE loss, (b) perceptual loss with VGG network~\cite{johnson2016perceptual}, (c) discriminator feature matching loss, and (d) combined with progressive reconstruction. }
	\label{loss}
	\vspace{-0.5cm}
\end{figure}

\begin{figure}[h!]
	\centering
	\vspace{-0.4cm}
	\includegraphics[width=\linewidth]{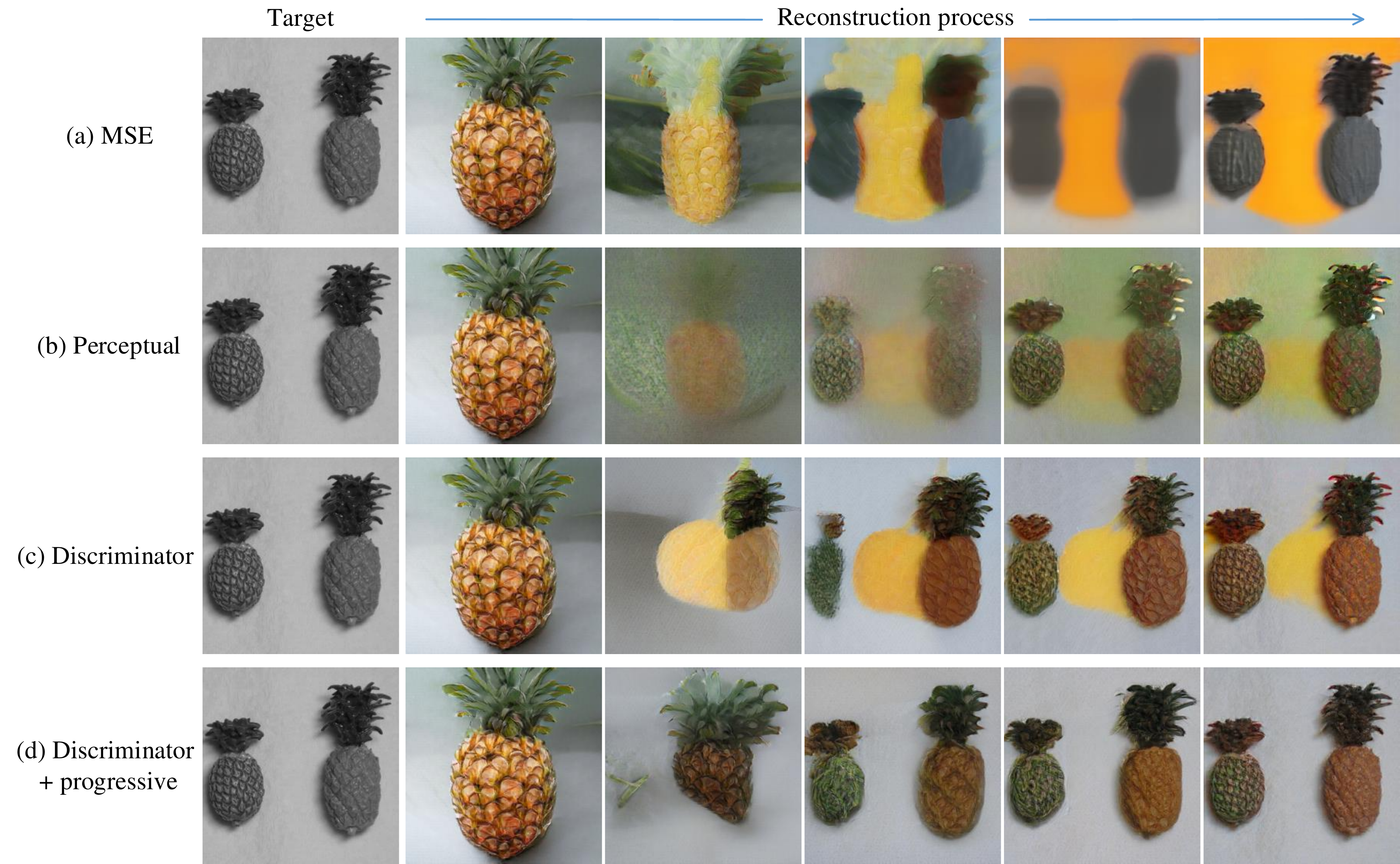}
	\vspace{-0.7cm}
	\caption{Comparison of different loss types and optimization techniques when fine-tuning the generator to restore the image.}
	\label{loss2}
	\vspace{-0.5cm}
\end{figure}

\begin{figure}[t!]
	\centering
	\includegraphics[width=\linewidth]{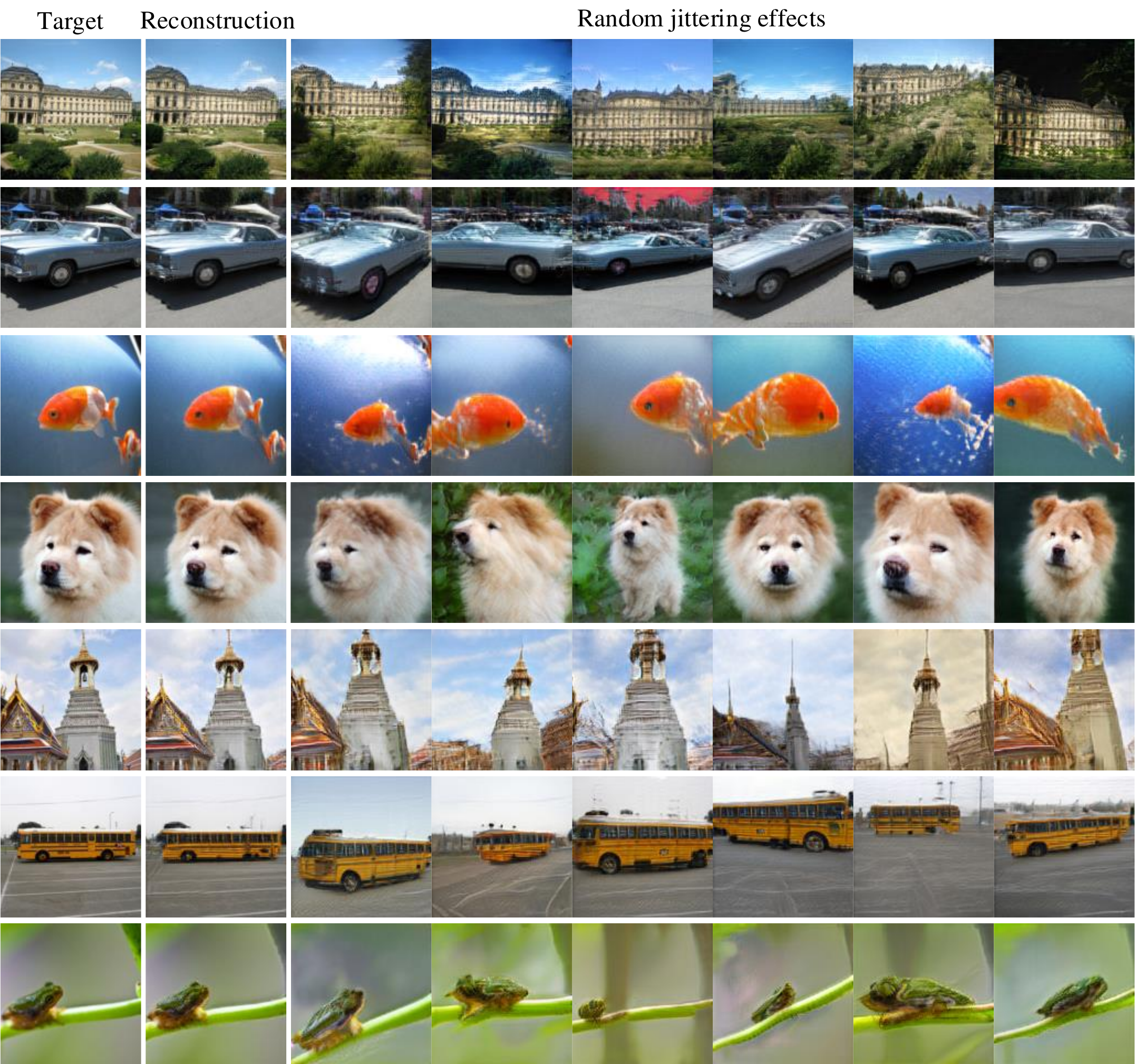}
	\caption{\textbf{Random jittering.} This is an extension of Fig.11 in the main paper. }
	\label{sup:jittering}
\end{figure}

\begin{figure}[t!]
	\centering
	\includegraphics[width=\linewidth]{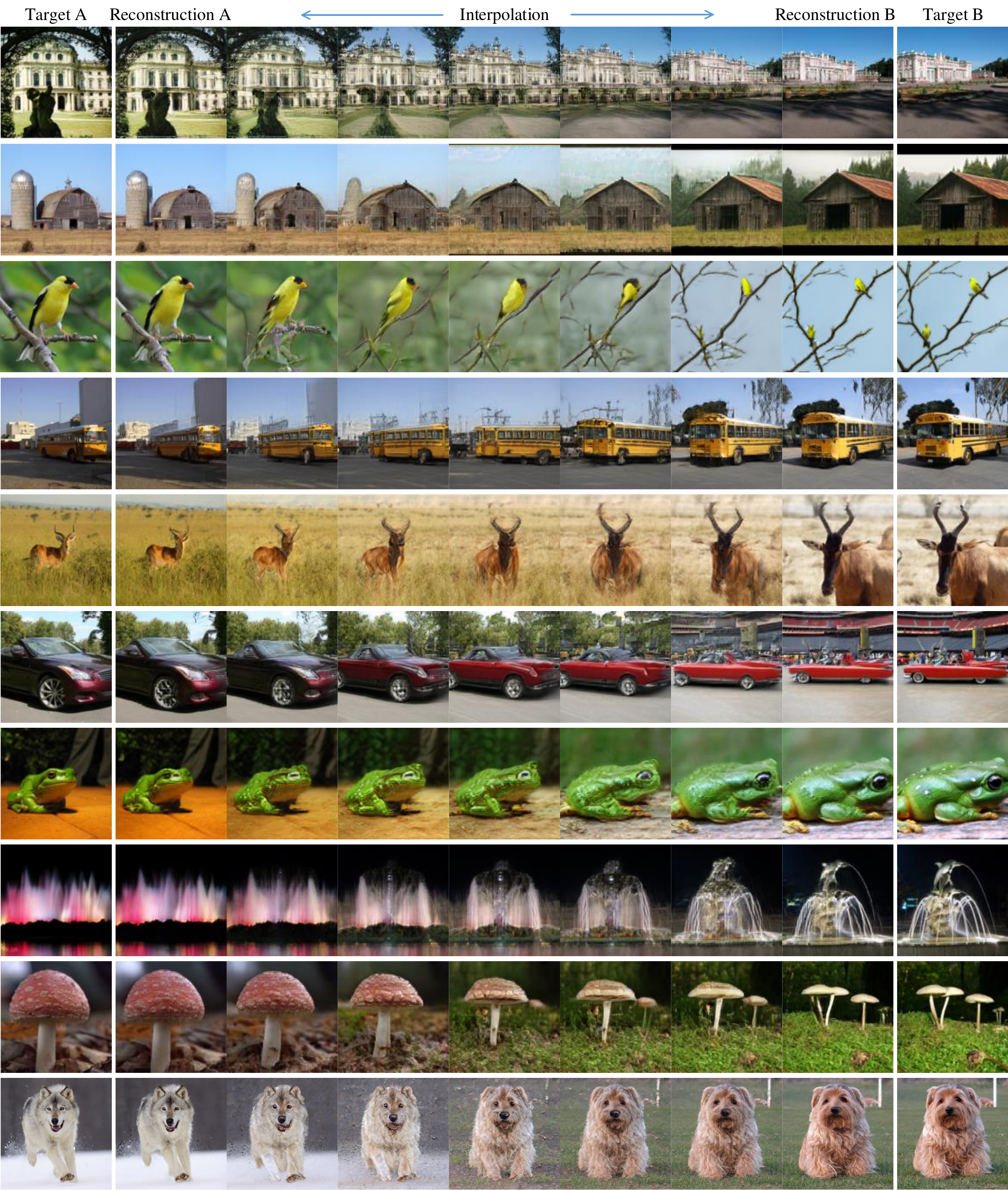}
	\caption{\textbf{Imaeg morphing.} This is an extension of Fig.12 in the main paper.}
	\label{sup:morphing}
\end{figure}

\begin{figure}[t!]
	\centering
	\includegraphics[width=\linewidth]{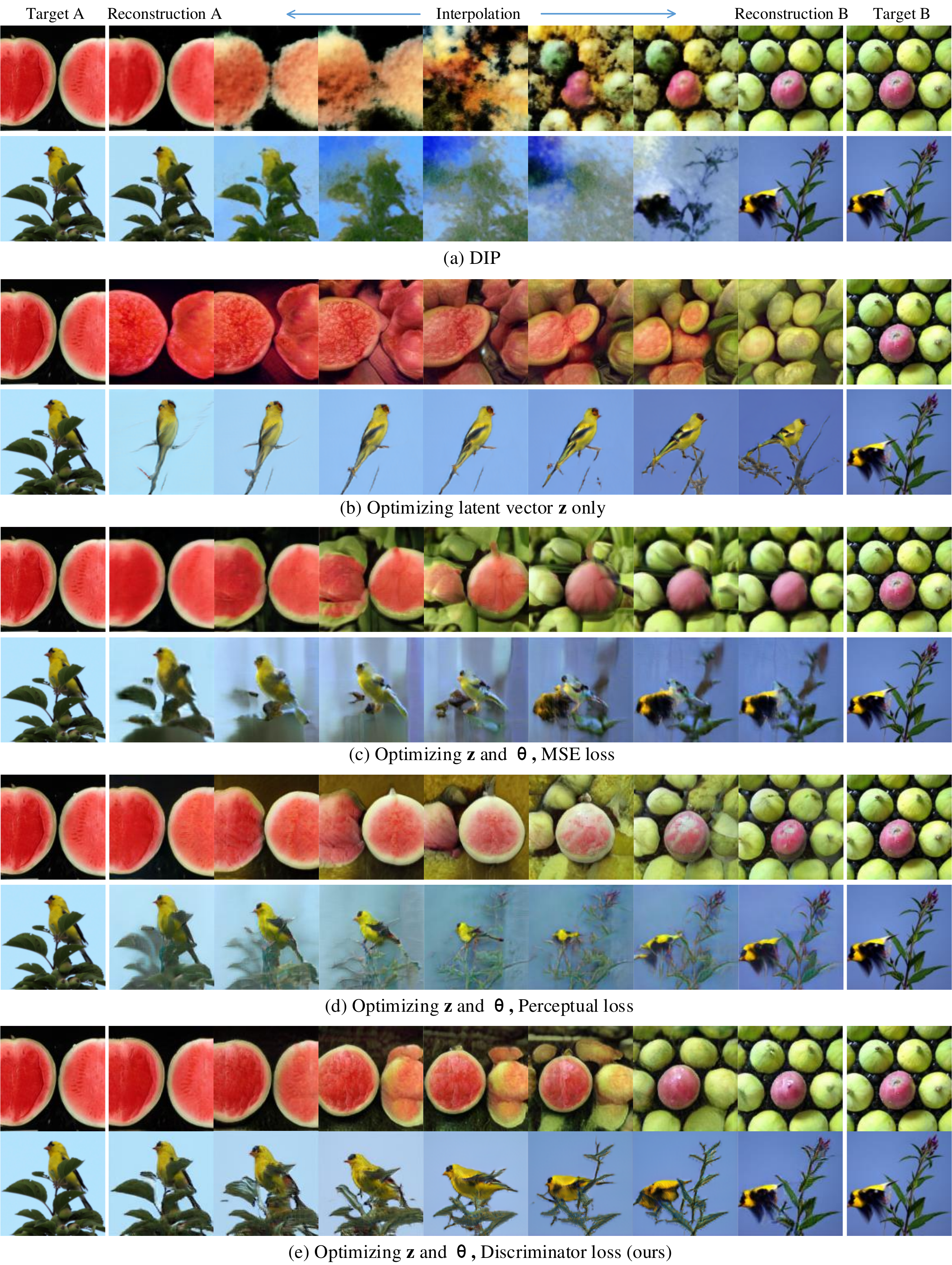}
	\caption{Comparison of various methods in image morphing, including (a) using DIP, (b) optimizing the latent vector \textbf{z} of the pre-trained GAN, and (c)(d)(e) optimizing both $\vz$ and the generator parameter $\vtheta$ with (c) MSE loss, (d) perceptual loss with VGG network~\cite{johnson2016perceptual}, and (e) discriminator feature matching loss.  (b) fails to produce accurate reconstruction while (a)(c)(d) could not obtain realistic interpolation results. In contrast, our results in (e) are much better. }
	\label{morphing_baseline}
\end{figure}

\begin{figure}[t!]
	\centering
	\includegraphics[width=\linewidth]{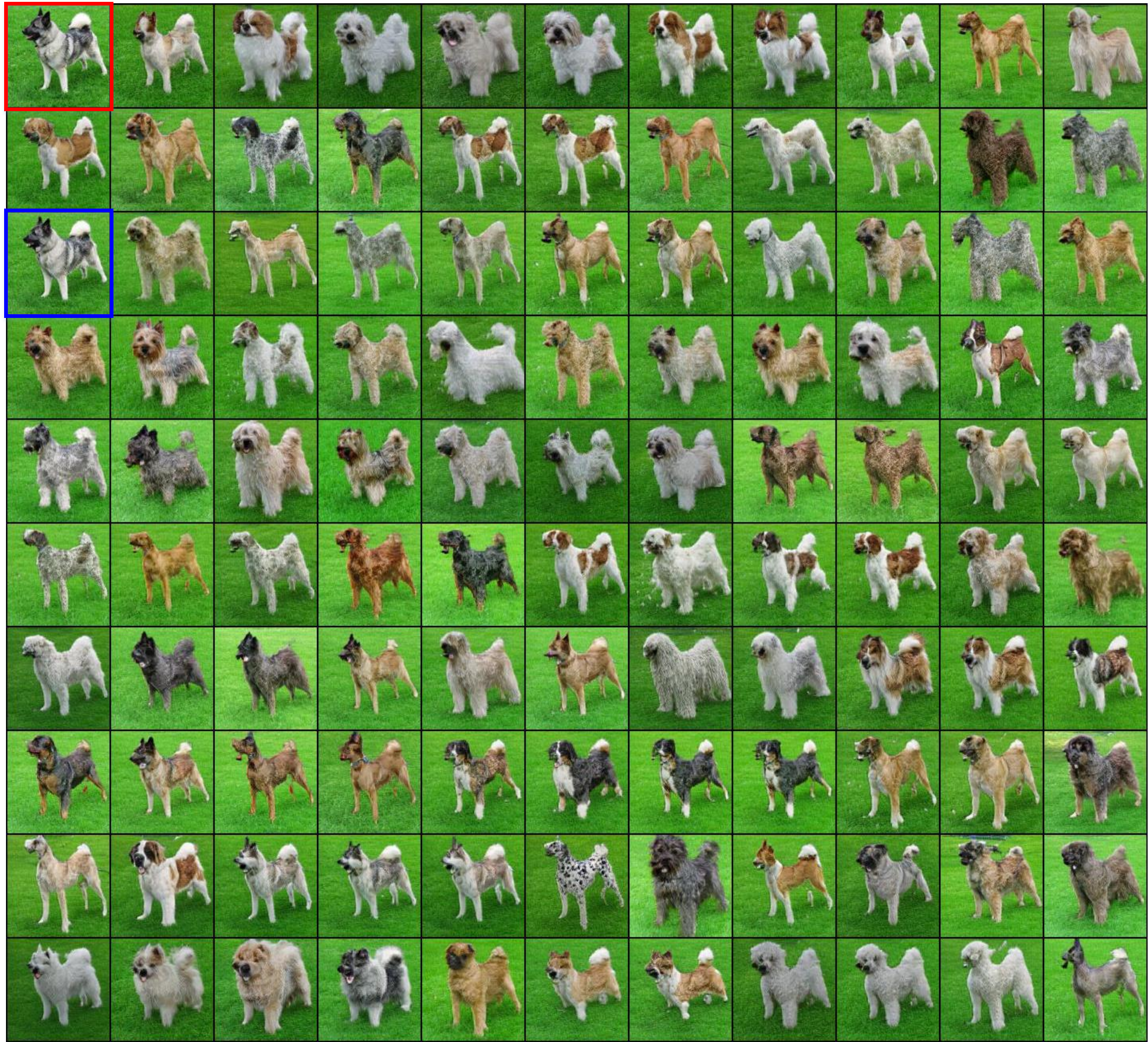}
	\caption{\textbf{Category transfer.} The \textcolor{red}{red} box shows the target, and the \textcolor{blue}{blue} box shows the reconstruction. Others are category transfer results.}
	\label{sup:category}
\end{figure}

We extend the figures of the main paper with more examples, as shown from Fig.~\ref{sup:colorization} to Fig.~\ref{sup:category}.

\clearpage

\clearpage

\clearpage

\section{Implementation Details}

\noindent\textbf{Architectures.}
We adopt the BigGAN\cite{brock2018large} architechtures of $128^2$ and $256^2$ resolutions in our experiments.
For the $128^2$ resolution, we use the best setting of \cite{brock2018large}, which has a channel multiplier of 96 and a batchsize of 2048.
As for the $256^2$ resolution, the channel multiplier and batchsize are respectively set to 64 and 1920 due to limited GPU resources.
We train the GANs on the ImageNet training set, and the $128^2$ and $256^2$ versions have Inception scores of 103.5 and 94.5 respectively.
Our experiments are conducted based on PyTorch~\cite{paszke2017automatic}.

\vspace{3pt}
\noindent\textbf{Initialization.}
In order to ease the optimization goal of Eq.4 in the paper, it is a good practice to start with a latent vector $\vz$ that produces an approximate reconstruction.
Therefore, we randomly sample 500 images using the GAN, and select the nearest neighbor of the target image under the discriminator feature metric as the starting point.
Since encoder based methods tend to fail for degraded input images, they are not used in this work.

Note that in BigGAN, a class condition is needed as input.
Therefore, in order to reconstruct an image, its class condition is required.
This image classification problem could be solved by training a corresponding deep network classfier and is not the focus of this work, hence we assume the class label is given except for the adversarial defense task.
For adversarial defense and images whose classes are not given, both the latent vector $\vz$ and the class condition are randomly sampled.

\vspace{3pt}
\noindent\textbf{Fine-tuning.}
With the above pre-trained BigGAN and initailized latent vector $\vz$, we fine-tune both the generator and the latent vector to reconstruct a target image.
As the batchsize is only 1 during fine-tuning, we use the tracked global statistics (\ie, running mean and running variance) for the batch normalization (BN)~\cite{ioffe2015batch} layers to prevent inaccurate statistic estimation.
The discriminator of BigGAN is composed of a number of residual blocks (6 blocks and 7 blocks for $128^2$ and $256^2$ resolution versions respectively).
The output features of these blocks are used as the discriminator loss, as described in Eq.(6) of the paper.
In order to prevent the latent vector from deviating too much from the prior gaussian distribution, we add an additional L2 loss to the latent vector $\vz$ with a loss weight of 0.02.
We adopt the ADAM optimizer~\cite{kingma2014adam} in all our experiments.
The detailed training settings for various tasks are listed from Table.\ref{table:color_setting} to Table.\ref{table:manipulation_setting}, where the parameters in these tables are explained below:

\vspace{5pt}
\noindent \textit{Blocks num.}: the number of generator blocks to be fine-tuned. For example, for blocks num.=1, only the shallowest block is fine-tuned. 

\noindent \textit{D loss weight}: the factor multiplied to the discriminator loss.

\noindent \textit{MSE loss weight}: the factor multiplied to the MSE loss.

\noindent \textit{Iterations}: number of training iterations of each stage.

\noindent \textit{G lr}: the learning rate of the generator blocks.

\noindent \textit{z lr}: the learning rate of the latent vector $\vz$.

\setlength{\tabcolsep}{4pt}
\begin{table}[!h]
	\parbox{.5\linewidth}{
		\centering
		\caption{The fine-tuning setting of colorization. The explanation of these parameters are in the main text}
		\resizebox{6.3cm}{1.35cm}{
			\begin{tabular}{l|ccccc}
				\Xhline{2\arrayrulewidth}
				Stage &  1 &  2 &  3 &  4 &  5 \\ \hline\hline
				Blocks num.  & 1       & 2       & 3       & 4    &  5 \\
				D loss weight   & 1       & 1       & 1       & 1   & 1    \\
				MSE loss weight & 0       & 0       & 0       & 0   & 0    \\
				Iterations      & 200     & 200     & 300     & 400  & 300   \\
				G lr            & 5e-5    & 5e-5    & 5e-5    & 5e-5  & 2e-5  \\
				z lr            & 2e-3    & 1e-3    & 5e-4    & 5e-5  & 2e-5  \\ \Xhline{2\arrayrulewidth}
			\end{tabular}
		}
		\label{table:color_setting}
	}
	\quad
	\parbox{.45\linewidth}{
		\centering
		\caption{The fine-tuning setting of inpainting. In this task we also fine-tune the class embedding apart from the generator blocks}
		\resizebox{5.6cm}{1.35cm}{
			\begin{tabular}{l|cccc}
				\Xhline{2\arrayrulewidth}
				Stage	&  1 &  2 &  3 &  4 \\ \hline\hline
				Blocks num.  & 5      & 5       & 5       & 5   \\
				D loss weight    & 1       & 1       & 0.1   & 0.1    \\
				MSE loss weigh   & 1       & 1       & 100   & 100     \\
				Iterations      & 400     & 200     & 200  &  200  \\
				G lr            & 2e-4    & 1e-4    & 1e-4  & 1e-5  \\
				z lr            & 1e-3    & 1e-4    & 1e-4  & 1e-5  \\ \Xhline{2\arrayrulewidth}
			\end{tabular}
		}
		\label{table:inpainting_setting}
	}
\end{table}
\setlength{\tabcolsep}{1.4pt}

\setlength{\tabcolsep}{4pt}
\begin{table}[!h]
	\parbox{.48\linewidth}{
		\centering
		\caption{The fine-tuning setting of super-resolution. This setting is biased towards MSE loss}
		\resizebox{6.1cm}{1.35cm}{
			\begin{tabular}{l|ccccc}
				\Xhline{2\arrayrulewidth}
				Stage &  1 &  2 &  3 &  4 &  5 \\ \hline\hline
				Blocks num.  & 1      & 2       & 3       & 4   & 5  \\
				D loss weight   & 1       & 1       & 1       & 0.5   & 0.1   \\
				MSE loss weight & 1       & 1       & 1       & 50   & 100    \\
				Iterations      & 200     & 200     & 200     & 200  &  200  \\
				G lr            & 2e-4    & 2e-4    & 1e-4    & 1e-4  & 1e-5  \\
				z lr            & 1e-3    & 1e-3    & 1e-4    & 1e-4  & 1e-5  \\ \Xhline{2\arrayrulewidth}
			\end{tabular}
		}
		\label{table:SR_setting_MSE}
	}
	\quad
	\parbox{.48\linewidth}{
		\centering
		\caption{The fine-tuning setting of super-resolution. This setting is biased towards discriminator loss}
		\resizebox{6.1cm}{1.35cm}{
			\begin{tabular}{l|ccccc}
				\Xhline{2\arrayrulewidth}
				Stage &  1 &  2 &  3 &  4 &  5 \\ \hline\hline
				Blocks num.  & 1      & 2       & 3       & 4   & 5  \\
				D loss weight   & 1       & 1       & 1       & 1   & 1    \\
				MSE loss weight & 1       & 1       & 1       & 1   & 1     \\
				Iterations      & 200     & 200     & 200     & 200  &  200  \\
				G lr            & 5e-5    & 5e-5    & 2e-5    & 1e-5  & 1e-5  \\
				z lr            & 2e-3    & 1e-3    & 2e-5    & 1e-5  & 1e-5  \\ \Xhline{2\arrayrulewidth}
			\end{tabular}
		}
		\label{table:SR_setting_D}
	}
\end{table}
\setlength{\tabcolsep}{1.4pt}

\begin{table}[!t]
	\parbox{.48\linewidth}{
		\centering
		\caption{The fine-tuning setting of adversarial defense. The fine-tuning is stopped if the MSE loss reaches 5e-3}
		\resizebox{4.2cm}{1.4cm}{
			\begin{tabular}{l|cccc}
				\Xhline{2\arrayrulewidth}
				& stage 1 & stage 2  \\ \hline\hline
				Blocks num.  & 6      & 6        \\
				D loss weight   & 0       & 0       \\
				MSE loss weight & 1       & 1       \\
				Iterations      & 100     & 900     \\
				G lr            & 2e-7    & 1e-4    \\
				z lr            & 5e-2    & 1e-4    \\ \Xhline{2\arrayrulewidth}
			\end{tabular}
		}
\label{table:defense}
	}
	\quad
	\parbox{.48\linewidth}{
		\centering
	\caption{The fine-tuning setting of manipulation tasks including random jittering, image morphing, and category transfer}
	\resizebox{5.2cm}{1.4cm}{
		\begin{tabular}{l|ccc}
			\Xhline{2\arrayrulewidth}
			& stage 1 & stage 2 & stage 3 \\ \hline\hline
			Blocks num.  & 5      & 5       & 5  \\
			D loss weight   & 1       & 1       & 1  \\
			MSE loss weight & 0       & 0       & 0    \\
			Iterations      & 125     & 125     & 100  \\
			G lr            & 2e-7    & 2e-5    & 2e-6  \\
			z lr            & 1e-1    & 2e-3    & 2e-6  \\ \Xhline{2\arrayrulewidth}
		\end{tabular}
	}
	\label{table:manipulation_setting}
	}
\end{table}

\vspace{3pt}
\noindent 
For inpainting and super-resolution, we use a weighted combination of discriminator loss and MSE loss, as the MSE loss is beneficial for the PSNR metric.
We also seamlessly replace BN with instance normalization (IN) for the setting in Table.~\ref{table:inpainting_setting}, Table.~\ref{table:SR_setting_MSE}, and Table.~\ref{table:defense}, which enables higher learning rate and leads to better PSNR.
This is achieved by initialize the scale and shift parameters of IN with the statistics of the output features of BN.
Our quantitative results on adversarial defense is based on the $256^2$ resolution model, while those for other tasks are based on the $128^2$ resolution models.

%
%

%% file: eccv2020submission.bbl
\begin{thebibliography}{10}
\providecommand{\url}[1]{\texttt{#1}}
\providecommand{\urlprefix}{URL }
\providecommand{\doi}[1]{https://doi.org/#1}

\bibitem{abdal2019image2stylegan}
Abdal, R., Qin, Y., Wonka, P.: Image2stylegan: How to embed images into the
  stylegan latent space? In: ICCV. pp. 4432--4441 (2019)

\bibitem{Albright2019CVPRWorkshops}
Albright, M., McCloskey, S.: Source generator attribution via inversion. In:
  CVPR Workshops (2019)

\bibitem{baluja2017adversarial}
Baluja, S., Fischer, I.: Adversarial transformation networks: Learning to
  generate adversarial examples. arXiv preprint arXiv:1703.09387  (2017)

\bibitem{bau2019semantic}
Bau, D., Strobelt, H., Peebles, W., Wulff, J., Zhou, B., Zhu, J.Y., Torralba,
  A.: Semantic photo manipulation with a generative image prior. ACM
  Transactions on Graphics (TOG)  \textbf{38}(4), ~59 (2019)

\bibitem{bau2019seeing}
Bau, D., Zhu, J.Y., Wulff, J., Peebles, W., Strobelt, H., Zhou, B., Torralba,
  A.: Seeing what a gan cannot generate. In: ICCV. pp. 4502--4511 (2019)

\bibitem{bigdeli2017deep}
Bigdeli, S.A., Zwicker, M., Favaro, P., Jin, M.: Deep mean-shift priors for
  image restoration. In: NIPS. pp. 763--772 (2017)

\bibitem{brock2018large}
Brock, A., Donahue, J., Simonyan, K.: Large scale gan training for high
  fidelity natural image synthesis. In: ICLR (2019)

\bibitem{chen2016trainable}
Chen, Y., Pock, T.: Trainable nonlinear reaction diffusion: A flexible
  framework for fast and effective image restoration. TPAMI  \textbf{39}(6),
  1256--1272 (2016)

\bibitem{choi2018stargan}
Choi, Y., Choi, M., Kim, M., Ha, J.W., Kim, S., Choo, J.: Stargan: Unified
  generative adversarial networks for multi-domain image-to-image translation.
  In: CVPR (2018)

\bibitem{creswell2018inverting}
Creswell, A., Bharath, A.A.: Inverting the generator of a generative
  adversarial network. In: IEEE transactions on neural networks and learning
  systems (2018)

\bibitem{dai2017towards}
Dai, B., Fidler, S., Urtasun, R., Lin, D.: Towards diverse and natural image
  descriptions via a conditional gan. In: ICCV. pp. 2970--2979 (2017)

\bibitem{deng2009imagenet}
Deng, J., Dong, W., Socher, R., Li, L.J., Li, K., Fei-Fei, L.: Imagenet: A
  large-scale hierarchical image database. In: CVPR. pp. 248--255 (2009)

\bibitem{donahue2017adversarial}
Donahue, J., Kr{\"a}henb{\"u}hl, P., Darrell, T.: Adversarial feature learning.
  In: ICLR (2017)

\bibitem{dong2015image}
Dong, C., Loy, C.C., He, K., Tang, X.: Image super-resolution using deep
  convolutional networks. In: TPAMI. vol.~38, pp. 295--307. IEEE (2015)

\bibitem{geman1984stochastic}
Geman, S., Geman, D.: Stochastic relaxation, gibbs distributions, and the
  bayesian restoration of images. TPAMI (6),  721--741 (1984)

\bibitem{goodfellow2014generative}
Goodfellow, I., Pouget-Abadie, J., Mirza, M., Xu, B., Warde-Farley, D., Ozair,
  S., Courville, A., Bengio, Y.: Generative adversarial nets. In: NIPS. pp.
  2672--2680 (2014)

\bibitem{gu2020image}
Gu, J., Shen, Y., Zhou, B.: Image processing using multi-code gan prior. CVPR
  (2020)

\bibitem{he2010single}
He, K., Sun, J., Tang, X.: Single image haze removal using dark channel prior.
  TPAMI  \textbf{33}(12),  2341--2353 (2010)

\bibitem{he2016deep}
He, K., Zhang, X., Ren, S., Sun, J.: Deep residual learning for image
  recognition. In: CVPR. pp. 770--778 (2016)

\bibitem{hussein2019image}
Hussein, S.A., Tirer, T., Giryes, R.: Image-adaptive gan based reconstruction.
  arXiv preprint arXiv:1906.05284  (2019)

\bibitem{ioffe2015batch}
Ioffe, S., Szegedy, C.: Batch normalization: Accelerating deep network training
  by reducing internal covariate shift. In: ICML. pp. 448--456 (2015)

\bibitem{johnson2016perceptual}
Johnson, J., Alahi, A., Fei-Fei, L.: Perceptual losses for real-time style
  transfer and super-resolution. In: ECCV. pp. 694--711. Springer (2016)

\bibitem{karras2019style}
Karras, T., Laine, S., Aila, T.: A style-based generator architecture for
  generative adversarial networks. In: CVPR. pp. 4401--4410 (2019)

\bibitem{kingma2014adam}
Kingma, D.P., Ba, J.: Adam: A method for stochastic optimization. arXiv
  preprint arXiv:1412.6980  (2014)

\bibitem{larsson2016learning}
Larsson, G., Maire, M., Shakhnarovich, G.: Learning representations for
  automatic colorization. In: ECCV. pp. 577--593. Springer (2016)

\bibitem{ledig2017photo}
Ledig, C., Theis, L., Husz{\'a}r, F., Caballero, J., Cunningham, A., Acosta,
  A., Aitken, A., Tejani, A., Totz, J., Wang, Z., et~al.: Photo-realistic
  single image super-resolution using a generative adversarial network. In:
  CVPR. pp. 4681--4690 (2017)

\bibitem{mittal2012making}
Mittal, A., Soundararajan, R., Bovik, A.C.: Making a “completely blind”
  image quality analyzer. IEEE Signal processing letters  \textbf{20}(3),
  209--212 (2012)

\bibitem{nguyen2015deep}
Nguyen, A., Yosinski, J., Clune, J.: Deep neural networks are easily fooled:
  High confidence predictions for unrecognizable images. In: CVPR. pp. 427--436
  (2015)

\bibitem{paszke2017automatic}
Paszke, A., Gross, S., Chintala, S., Chanan, G., Yang, E., DeVito, Z., Lin, Z.,
  Desmaison, A., Antiga, L., Lerer, A.: Automatic differentiation in pytorch
  (2017)

\bibitem{roth2005fields}
Roth, S., Black, M.J.: Fields of experts: A framework for learning image
  priors. In: CVPR. pp. 860--867 (2005)

\bibitem{rudin1992nonlinear}
Rudin, L.I., Osher, S., Fatemi, E.: Nonlinear total variation based noise
  removal algorithms. Physica D: nonlinear phenomena  \textbf{60}(1-4),
  259--268 (1992)

\bibitem{salimans2016improved}
Salimans, T., Goodfellow, I., Zaremba, W., Cheung, V., Radford, A., Chen, X.:
  Improved techniques for training gans. In: NIPS. pp. 2234--2242 (2016)

\bibitem{samangouei2018defense}
Samangouei, P., Kabkab, M., Chellappa, R.: Defense-gan: Protecting classifiers
  against adversarial attacks using generative models. In: ICLR (2018)

\bibitem{shaham2019singan}
Shaham, T.R., Dekel, T., Michaeli, T.: Singan: Learning a generative model from
  a single natural image. In: ICCV. pp. 4570--4580 (2019)

\bibitem{shen2019interpreting}
Shen, Y., Gu, J., Tang, X., Zhou, B.: Interpreting the latent space of gans for
  semantic face editing. CVPR  (2020)

\bibitem{ulyanov2018deep}
Ulyanov, D., Vedaldi, A., Lempitsky, V.: Deep image prior. In: CVPR. pp.
  9446--9454 (2018)

\bibitem{wang2018high}
Wang, T.C., Liu, M.Y., Zhu, J.Y., Tao, A., Kautz, J., Catanzaro, B.:
  High-resolution image synthesis and semantic manipulation with conditional
  gans. In: CVPR (2018)

\bibitem{wang2019deep}
Wang, X., Yu, K., Dong, C., Tang, X., Loy, C.C.: Deep network interpolation for
  continuous imagery effect transition. In: CVPR. pp. 1692--1701 (2019)

\bibitem{xiangli2020real}
Xiangli*, Y., Deng*, Y., Dai*, B., Loy, C.C., Lin, D.: Real or not real, that
  is the question. In: ICLR (2020)

\bibitem{yang2019semantic}
Yang, C., Shen, Y., Zhou, B.: Semantic hierarchy emerges in deep generative
  representations for scene synthesis. arXiv preprint arXiv:1911.09267  (2019)

\bibitem{yeh2017semantic}
Yeh, R.A., Chen, C., Yian~Lim, T., Schwing, A.G., Hasegawa-Johnson, M., Do,
  M.N.: Semantic image inpainting with deep generative models. In: CVPR. pp.
  5485--5493 (2017)

\bibitem{zhang2017learning}
Zhang, K., Zuo, W., Gu, S., Zhang, L.: Learning deep cnn denoiser prior for
  image restoration. In: CVPR. pp. 3929--3938 (2017)

\bibitem{zhang2016colorful}
Zhang, R., Isola, P., Efros, A.A.: Colorful image colorization. In: ECCV. pp.
  649--666. Springer (2016)

\bibitem{zhou2017places}
Zhou, B., Lapedriza, A., Khosla, A., Oliva, A., Torralba, A.: Places: A 10
  million image database for scene recognition. TPAMI  \textbf{40},  1452--1464
  (2017)

\bibitem{zhu2016generative}
Zhu, J.Y., Kr{\"a}henb{\"u}hl, P., Shechtman, E., Efros, A.A.: Generative
  visual manipulation on the natural image manifold. In: ECCV (2016)

\bibitem{zhu2017unpaired}
Zhu, J.Y., Park, T., Isola, P., Efros, A.A.: Unpaired image-to-image
  translation using cycle-consistent adversarial networks. In: ICCV (2017)

\bibitem{zhu1997prior}
Zhu, S.C., Mumford, D.: Prior learning and gibbs reaction-diffusion. TPAMI
  \textbf{19}(11),  1236--1250 (1997)

\end{thebibliography}
